\documentclass[]{aa}

\usepackage{rotating}
\usepackage{graphicx}
\usepackage{sidecap}
\usepackage{subcaption}
\usepackage{url}
\usepackage[colorlinks=true]{hyperref}
\usepackage{soul}
\hypersetup{allcolors=blue}
\usepackage{microtype} 

\usepackage{txfonts}
\usepackage[]{xcolor}
\usepackage{float}

\bibpunct{(}{)}{;}{a}{}{,}
\DeclareUnicodeCharacter{00A0}{~}
\usepackage[us,12hr]{datetime}
\usepackage{ulem}

\newcommand{\zbf}{\mathbf{z}}
\newcommand{\xbf}{\mathbf{x}}

\newcommand{\fbf}{\mathbf{f}}
\newcommand{\vbf}{\mathbf{v}}

\newcommand{\kms}{\,km\,s$^{-1}$}

\usepackage{cancel}
\usepackage{color}

\definecolor{deepmagenta}{rgb}{0.8, 0.0, 0.8}
\definecolor{green}{rgb}{0.0, 0.5, 0.1}

\def\fix#1{{#1}}
\def\fixii#1{{#1}}

\renewcommand{\jobname}{inspectorch}
\newcommand{\Inspectorch}{\texttt{Inspectorch}}

\begin{document}

\title{\Inspectorch: Efficient rare event exploration in solar observations}

\author{C. J. D\'iaz Baso
\inst{1,2}
\and
I. J. Soler Poquet
\inst{1,2}
\and
C. Kuckein
\inst{3,4}
\and
M. van Noort
\inst{5}
\and
N. Poirier 
\inst{6}
}

\institute{
Institute of Theoretical Astrophysics,
University of Oslo, %
P.O. Box 1029 Blindern, N-0315 Oslo, Norway
\and
Rosseland Centre for Solar Physics,
University of Oslo, %
P.O. Box 1029 Blindern, N-0315 Oslo, Norway
\and Instituto de Astrof\'isica de Canarias, C/V\'{\i}a L\'actea s/n, E-38205 La Laguna, Tenerife, Spain
\and Departamento de Astrof\'isica, Universidad de La Laguna, E-38206 La Laguna, Tenerife, Spain
\and Max-Planck Institute for Solar System Research, Justus-von-Liebig-Weg 3, 37077 G\"ottingen, Germany
\and LPC2E, OSUC, Univ Orl\'eans, CNRS, CNES, F-45071 Orl\'eans, France
\\
\email{carlos.diaz@astro.uio.no}
}

\date{Draft: compiled on \today\ at \currenttime~UT}

\authorrunning{D\'iaz Baso et al.}

\abstract
{\fix{Modern solar observatories resolve the solar atmosphere in unprecedented detail, enabling studies of its activity on spatial scales below 0.1~arcsec and temporal cadences of a few seconds.} However, the large volume of data collected by our telescopes cannot be fully analyzed with conventional methods. Popular machine learning methods identify general trends from observations, but tend to overlook unusual events due to their low frequency of occurrence.}
{Our aim in this work is to study the applicability of unsupervised probabilistic methods to efficiently identify rare events in multidimensional solar observations and optimize our computational resources to the study of these extreme phenomena.}
{To achieve this, we introduce \Inspectorch, an open-source framework that utilizes flow-based models—flexible density estimators capable of learning the multidimensional distribution of \fix{high-dimensional datasets}. Once optimized, it assigns a probability to each sample, allowing us to identify unusual events. We illustrate the potential of this approach by applying it to observations from the Hinode Spectro-Polarimeter (Hinode/SP), the Interface Region Imaging Spectrograph (IRIS), the Microlensed Hyperspectral Imager (MiHI) at Swedish 1-m Solar Telescope (SST), the Atmospheric Imaging Assembly (AIA) on board the Solar Dynamics Observatory (SDO) and the Extreme Ultraviolet Imager (EUI) on board Solar Orbiter.}
{On those datasets, we find that the algorithm assigns consistently lower probabilities to spectra that exhibit unusual features. For example, it identifies profiles with very strong Doppler shifts, uncommon broadening, complex profiles where the magnetic field polarity changes with height, and temporal dynamics associated with small-scale reconnection events, among others.}
{As a result, \Inspectorch\ demonstrates that density estimation using flow-based models offers a powerful approach to identifying rare events in large solar datasets. We find that the framework is robust across different data regimes and requires limited parameter tuning. The resulting probabilistic anomaly scores allow computational resources to be focused on the most informative and physically relevant events. We make our Python package publicly available at \url{https://github.com/cdiazbas/inspectorch}.}

\keywords{Sun: atmosphere -- Line: formation  -- Methods: data analysis -- Sun: activity -- Radiative transfer}

\maketitle



\section{Introduction}\label{sec:intro}

\begin{figure*}[t]
    \sidecaption
    \includegraphics[width=0.69\linewidth]{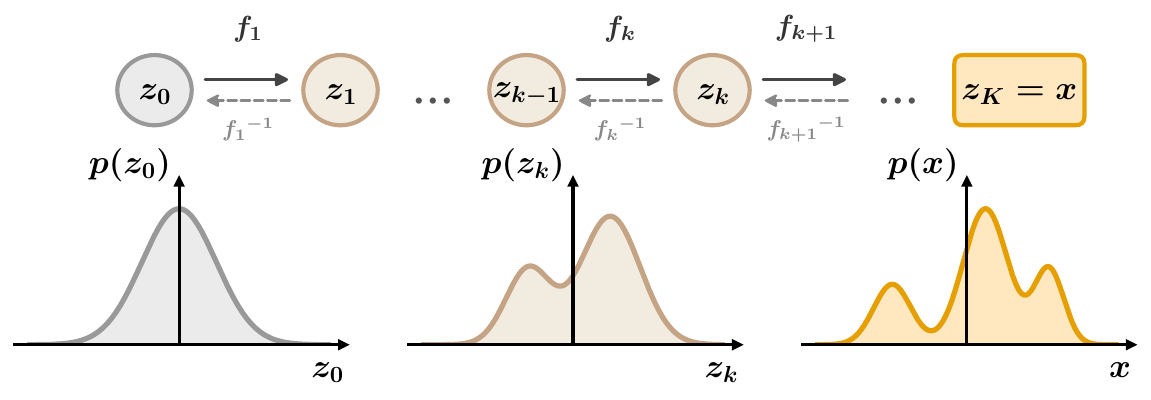}
    \caption{Schematic illustration of the Normalizing Flow approach for rare event detection for one-dimensional data. The model learns an invertible transformation that maps a simple distribution (typically a Gaussian) to a complex distribution. Samples can be generated by mapping from latent space to data space (solid arrows), while likelihoods are evaluated by the inverse mapping (dashed arrows). Events with very low probabilities correspond to rare or anomalous samples.}
\label{fig:nflows_sketch}
\end{figure*}

\fix{Modern and upcoming solar observational facilities are generating data at an ever-increasing rate, providing views of the Sun with unprecedented detail—reaching spatial scales below 0.1 arcsec, temporal cadences of a few seconds, and spectral resolving powers exceeding $R \sim 10^5$.} 
This dramatic increase in data volume poses major challenges for traditional analysis methods. Manual inspection of such vast datasets is impractical and risks overlooking rare events. Because these events occur infrequently, they remain among the least understood phenomena in solar physics, yet they are often central to many studies, as they may provide unique insights into the fundamental processes driving solar activity \citep[e.g.,][]{2011SSRv..159...19F,2014LRSP...11....3C,2019LRSP...16....1B}.

Some solar features are relatively easy to identify because of their clear signatures, whereas others are much harder to detect, as they may only manifest in specific wavelength ranges or over short temporal windows \citep[e.g.,][]{Bhatnagar2024,SolerPoquet2025}. A common strategy to gain physical insight into the underlying solar phenomena is to infer the physical parameters of the solar atmosphere (e.g., temperature, velocity, magnetic field) through spectropolarimetric inversions. However, these methods \citep[e.g.,][]{RuizCobo1992,delaCruz2019_STiC, 2022ApJ...933..145L} are generally too computationally expensive to be applied systematically to large datasets or extended time series, which limits their use in the context of big data. While recent efforts have successfully accelerated spectropolarimetric inversions \citep[e.g.,][]{Asensio2019,SainzDalda2019,DiazBaso2022_nflows}, this approach merely shifts the computational bottleneck from managing large raw datasets to the subsequent analysis and interpretation of massive volumes of inversion results.

An alternative approach is to analyze the spectra directly in a statistical manner, searching for patterns without relying on physical models or prior assumptions about the underlying processes (unsupervised learning). Clustering algorithms, for instance, can group similar sets of features (e.g., spectral lines, light curves, etc.) together and thereby reveal underlying structures in the data. Several successful applications of such methods exist in solar physics, \fix{pioneered by \citet{2007ApJ...663.1386P} and \citet{2011A&A...530A..14V}}. For example, \cite{Panos2018_flarekmeans} and \cite{2020A&A...640A..71K} employed k-means clustering to identify different types of spectral profiles (and hence atmospheric conditions) during flares and filament eruptions, demonstrating the potential of this technique to categorize solar events based solely on their spectral characteristics. 
%
\fix{However, because global-trend methods are optimized to capture the bulk variance of the data, isolating subtle rare events—beyond just extreme outliers—often requires impractically large cluster numbers or extensive tuning.}
For example, \citet{Bhatnagar2024} reported substantial difficulties when applying k-means to identify Quiet-Sun Ellerman Bombs (QSEBs). The rarity of QSEB spectral profiles forced them to use a very large number of clusters and careful hyperparameter tuning, and even then, a biased training was required to ensure that QSEB-like profiles were well represented. This illustrates the intrinsic challenge of using clustering techniques to capture unusual solar phenomena.

In contrast, anomaly detection offers a targeted approach to identify samples that deviate significantly from the norm \citep[e.g.,][]{Chandola2009,2021A&C....3600481L,2022FrASS...937863W,2023MNRAS.526.3072B}. While there are many strategies (reconstruction-based methods, clustering-based techniques, etc.), density-based methods are particularly powerful for scientific discovery: by modeling the full multidimensional probability distribution of the data, we can statistically define "rareness" as low probability. This allows us to quantify how anomalous a given spectrum is without defining a priori what we are looking for \citep[see the review by][for a detailed description]{Pang2021}.

While this task is straightforward in one dimension (essentially equivalent to constructing a histogram), it becomes much more challenging when dealing with high-dimensional data that combine many correlated wavelength measurements across several spectral windows, together with the full description of light and polarization through the Stokes parameters, as well as their spatial and temporal extension. To tackle this challenge, we employ Normalizing Flows \citep[NFs;][]{Papamakarios2019_review}. Unlike traditional methods, NFs can learn complex, high-dimensional distributions with high precision, making them well suited for the systematic discovery of rare events that may otherwise be missed \citep[e.g.,][]{Ciuca2022mla}. Building on their successful application to spectropolarimetric inference in solar physics \citep{DiazBaso2022_nflows}, we adapt NFs here to the complementary problem of unsupervised anomaly detection, where the goal is to characterize the distribution of observed spectra rather than inferring atmospheric parameters.

In this work, we present \Inspectorch, an open-source Python package that implements these powerful density estimators to enable the systematic discovery of rare events in solar spectropolarimetric observations. The paper is organized as follows. \fix{We demonstrate how Normalizing Flows serve} a robust baseline for unsupervised anomaly detection in solar physics, capable of assigning precise probabilities to complex spectra (Sect.~\ref{sec:problem_formulation}). Later, we demonstrate the method's effectiveness using data from diverse facilities (Hinode, IRIS, SST, SDO, Solar Orbiter), 
\fix{identifying peculiar data samples associated with uncommon, and likely interesting, physical phenomena} that standard methods overlook (Sect.~\ref{sec:results}).
Furthermore, we show how the framework naturally generalizes to detect anomalies in spatial and temporal domains. Finally, we explore Flow Matching as a preliminary future direction for scaling to even larger datasets (Sect.~\ref{sec:flow_matching}) and conclude with a brief discussion on the implications of this work (Sect.~\ref{sec:conclusions_summary}).

\section{Anomaly detection as density estimation}\label{sec:problem_formulation}

In solar physics, we are often interested in identifying events that are physically distinct from the dominant background. Statistically, if the observed data $\xbf$ (e.g., a spectral vector) are treated as samples drawn from an unknown probability distribution $p(\xbf)$, then common physical states correspond to regions of high probability density, while rare or anomalous states populate the low-density tails. Formally, we define an anomaly as any sample $\xbf$ satisfying $p(\xbf) < \epsilon$, for a user-defined threshold $\epsilon$. The task of unsupervised anomaly detection thus reduces to estimating the scalar function $p(\xbf)$ directly from the observed data, without labeled examples or prior assumptions about the underlying physical mechanisms.


\subsection{Normalizing Flows}\label{sec:nflows}

Classical density estimation techniques, such as Gaussian mixture models, struggle with the high dimensionality and non-linear correlations of modern solar observations. To overcome this, the field has increasingly turned to deep generative models—advanced neural networks designed to learn the underlying probability distribution of complex datasets. However, different generative approaches have important limitations for anomaly detection. Variational Autoencoders (VAEs), for example, rely on probabilistic latent representations and approximate optimization objectives, which limit their sensitivity to outliers. Generative Adversarial Networks (GANs), while capable of producing highly realistic synthetic data, do not provide an explicit probability for a given observation. Normalizing Flows offer a principled alternative: they are expressive enough to capture complex, multi-modal solar distributions, while mathematically guaranteeing exact likelihood calculations. Although capable of generating synthetic data, we exclusively exploit their density estimation properties here.

Normalizing Flows \citep[see e.g.,][for reviews]{Papamakarios2019_review, Kobyzev2019Review} model a complex data distribution $p_X(\xbf)$ by transforming a simple base distribution $p_Z(\zbf)$ (typically a standard multivariate Gaussian) through an invertible and differentiable mapping $\xbf =\fbf (\zbf)$. As illustrated in Fig.~\ref{fig:nflows_sketch}, for a one-dimensional example, samples are generated by mapping latent variables from the base distribution to the data space ($\zbf \rightarrow \xbf$), while likelihood evaluation requires the inverse mapping from data space back to latent space ($\xbf \rightarrow \zbf$).

Invertibility is therefore essential: the same transformation must support both sampling and exact probability evaluation. Differentiability ensures that local expansions and contractions of the density—visible in Fig.~\ref{fig:nflows_sketch}—are preserving the total probability.
In one dimension, for $x=f(z)$, probability conservation implies $p_X(x)\,dx = p_Z(z)\,dz$, hence $p_X(x)=p_Z(z)\left|\frac{dz}{dx}\right|$. The multidimensional case replaces the derivative by the Jacobian determinant,
yielding the following expression:

\begin{align}\label{eq:chng_vrbl}
p_X(\xbf) = p_Z(\zbf) \left| \det \left( \frac{\partial \zbf}{\partial \xbf} \right) \right|
&= p_Z(\zbf) \left| \det \left( \frac{\partial \fbf(\zbf)}{\partial \zbf} \right)^{-1} \right| \\ &= p_Z(\fbf^{-1}(\xbf)) \left| \det \left( \frac{\partial \fbf^{-1}(\xbf)}{\partial \xbf} \right) \right| 
\end{align}
where, in each of these expressions, the first factor represents the probability density for the base distribution evaluated at the corresponding point in the latent space and the second factor accounts for the change in the volume due to the transformation \fix{(taking the absolute value of the determinant to ensure the probability density remains strictly positive)}, forcing the total integrated probability to be unity.

If the overall transformation is obtained by composing a sequence of $K$ simple transformations, we can view this as a discrete evolution where $\zbf_0$ is the base distribution and $\zbf_K = \xbf$ the data distribution. The mapping is then expressed as $\xbf = \fbf(\zbf) = \fbf_K \circ \dots \circ \fbf_1(\zbf)$, their inverse can also be decomposed into the components {$\mathbf{f^{-1}}=(\mathbf{f}^{-1}_{K}\circ\cdots\circ \fbf^{-1}_{2}\circ \fbf^{-1}_{1})$} and the Jacobian determinant is the product of the determinant of each component. Therefore, the logarithm (which is often more numerically stable) of the probability of the overall transformation is then
\begin{equation}\label{eq:chng_vrbl_log}
\log p_X(\xbf) = \log p_Z(\mathbf{z}_0) + \sum^K_{k=1}\log \left|\det \left( \frac{\partial  \fbf_k (\mathbf{z}_{k-1})}{\partial {\mathbf{z}_{k-1}}} \right)^{-1} \right|
\end{equation}
where $\mathbf{z}_k=\mathbf{f}_k(\mathbf{z}_{k-1})$. 

The term Normalizing Flow refers to the compositional character described above and sketched in Fig.~\ref{fig:nflows_sketch}, where multiple transformations are iteratively changing the overall shape of the \fix{intermediate distributions $p_k(\mathbf{z}_k)$}.
This compositional perspective has been extended in more recent generative frameworks, including diffusion models and continuous-time flows trained via flow matching \citep{Chen2018, Ho2020, Song2021, Lipman2022}, which we briefly explore in Sect.~\ref{sec:flow_matching} as a future scaling direction.

Normalizing Flows are trained by minimizing the negative log-likelihood (NLL) of the observed data. For a dataset of $N$ samples $\{\mathbf{x}_i\}_{i=1}^N$, the objective function is:
\begin{equation}
    \mathcal{L}_{\text{NF}}(\phi) = -\frac{1}{N}\sum^N_{i=1} \log p_\phi(\mathbf{x}_i)
    ,
\end{equation}
where $\log p_\phi(\mathbf{x})$ is computed using Eq.~\ref{eq:chng_vrbl_log}. By minimizing $\mathcal{L}_{\text{NF}}$, the parameters $\phi$ of all $\fbf_k$ components are adjusted so that the transformed data matches the chosen base distribution, effectively mapping the highly correlated solar spectra into an uncorrelated multivariate Gaussian.

After training, the model assigns high likelihoods to common spectra and low likelihoods to rare or anomalous ones, allowing events to be ranked directly by their probability. As the method is fully unsupervised, low-probability spectra may correspond either to rare physical phenomena or to instrumental artifacts, both of which are valuable to identify.

\begin{figure*}[t]
    \centering
    \includegraphics[width=0.95\linewidth]{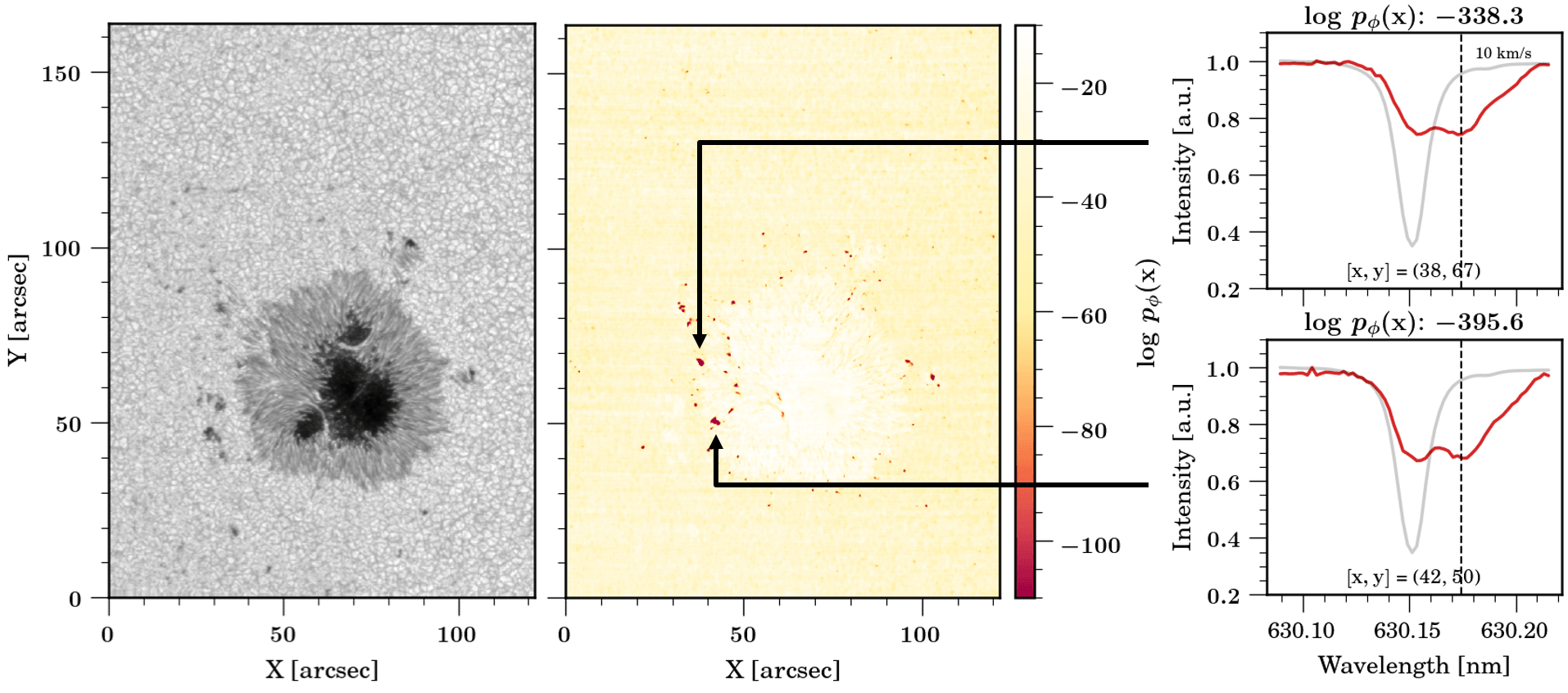}
    \caption{Application of \Inspectorch\ to a Hinode/SP dataset containing a sunspot. The left panels show the continuum intensity map at 630.10~nm and log-probability $\log p_\phi(\xbf)$ map. The color scale is clipped to highlight the most extreme values, although some pixels have log-probabilities below $-110$. The right panel shows an example of two of the most unusual spectra extracted from the low-probability regions. \fix{All spectral profiles are normalized to their own maximum value to facilitate shape comparison (a.u.: arbitrary units); the likelihood evaluation is performed on the unnormalized spectra.} A vertical dashed line indicates Doppler shifts of $10$~\kms and an average quiet-Sun profile in gray is shown for reference.}
    \label{fig:hinode_logprob}
\end{figure*}

\begin{figure*}[t]
    \sidecaption
    \begin{minipage}{0.68\linewidth}
        \includegraphics[width=0.91\linewidth, trim={0 40px 0 0},clip]{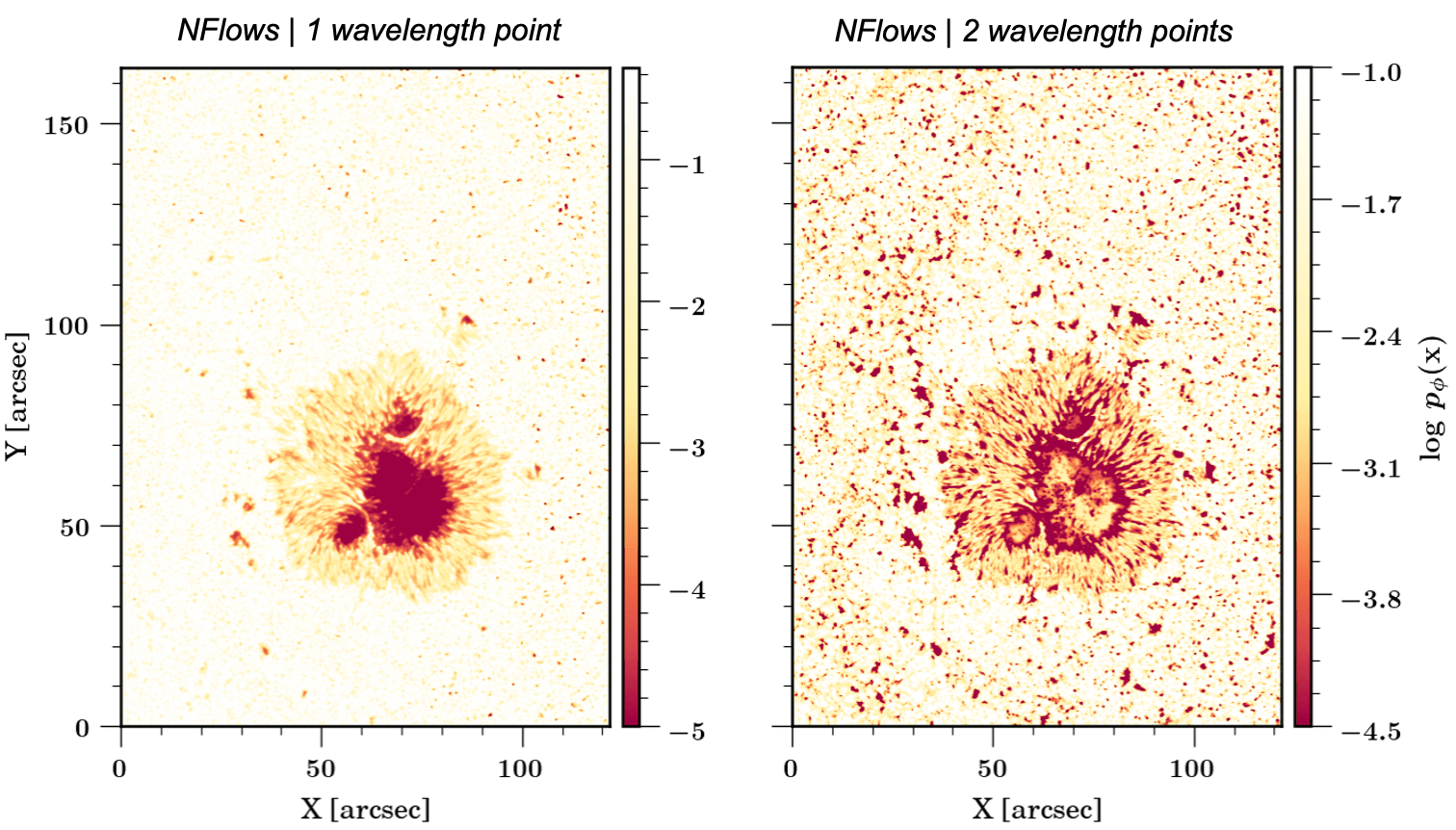}
        \includegraphics[width=0.91\linewidth]{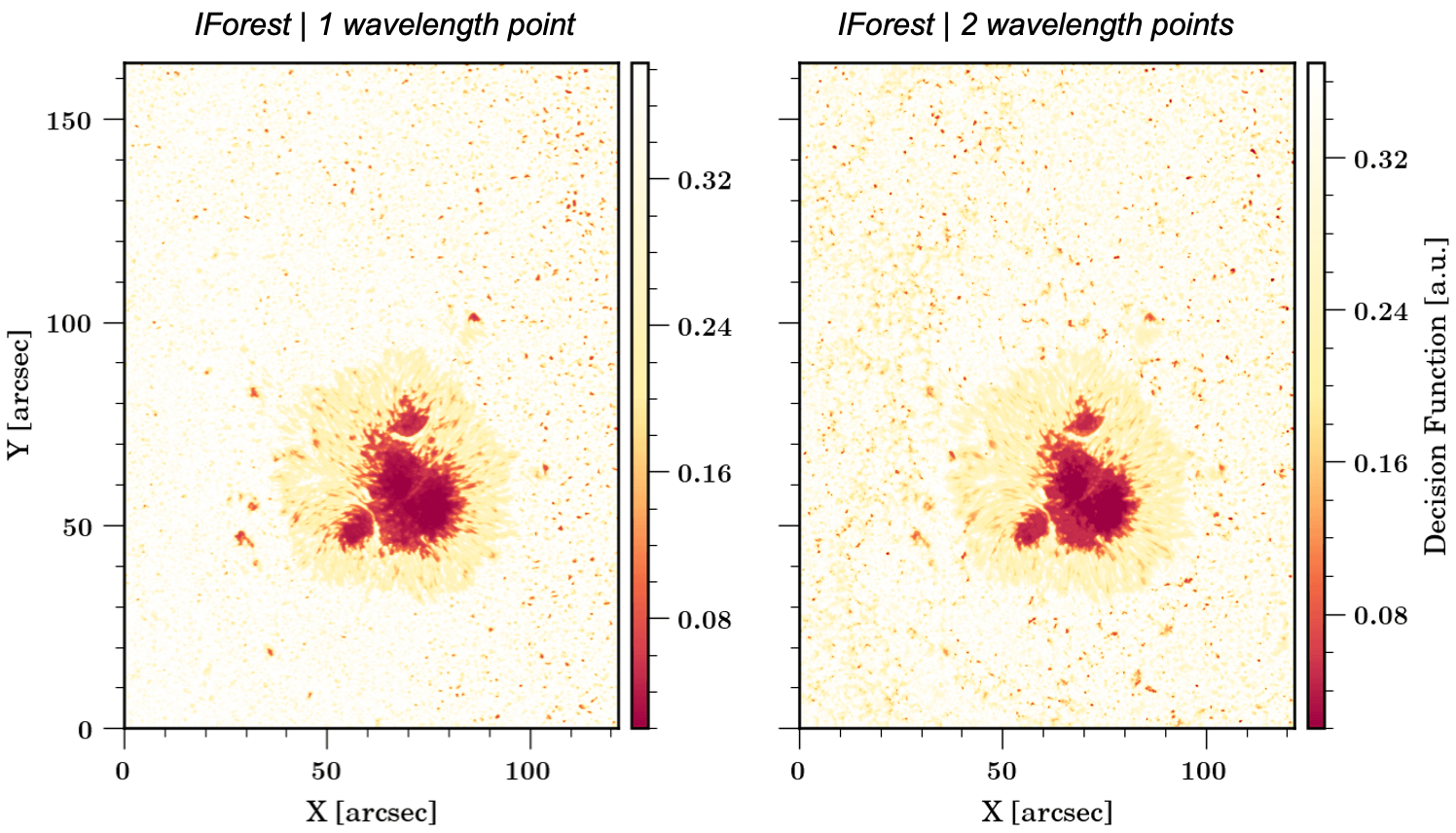}
    \end{minipage}
    \caption{Comparison between the application of Normalizing Flows and Isolation Forest to the same Hinode dataset shown in Fig.~\ref{fig:hinode_logprob}. The top panels show the log-probability map computed when using the method based on Normalizing Flows with the Hinode dataset using one wavelength point (the continuum at 6301.0~nm) on the left and two wavelength points (the continuum at 6301.0~nm and the core at 6301.5~nm) on the right. The bottom panels show the anomaly score map computed using the Isolation Forest method with the same dataset and same wavelength points. Note that Isolation Forest produces anomaly scores, not probabilities.}
    \label{fig:hinode_comparison_nflows}
\end{figure*}

\begin{figure*}[t]
    \centering
    \includegraphics[width=\linewidth]{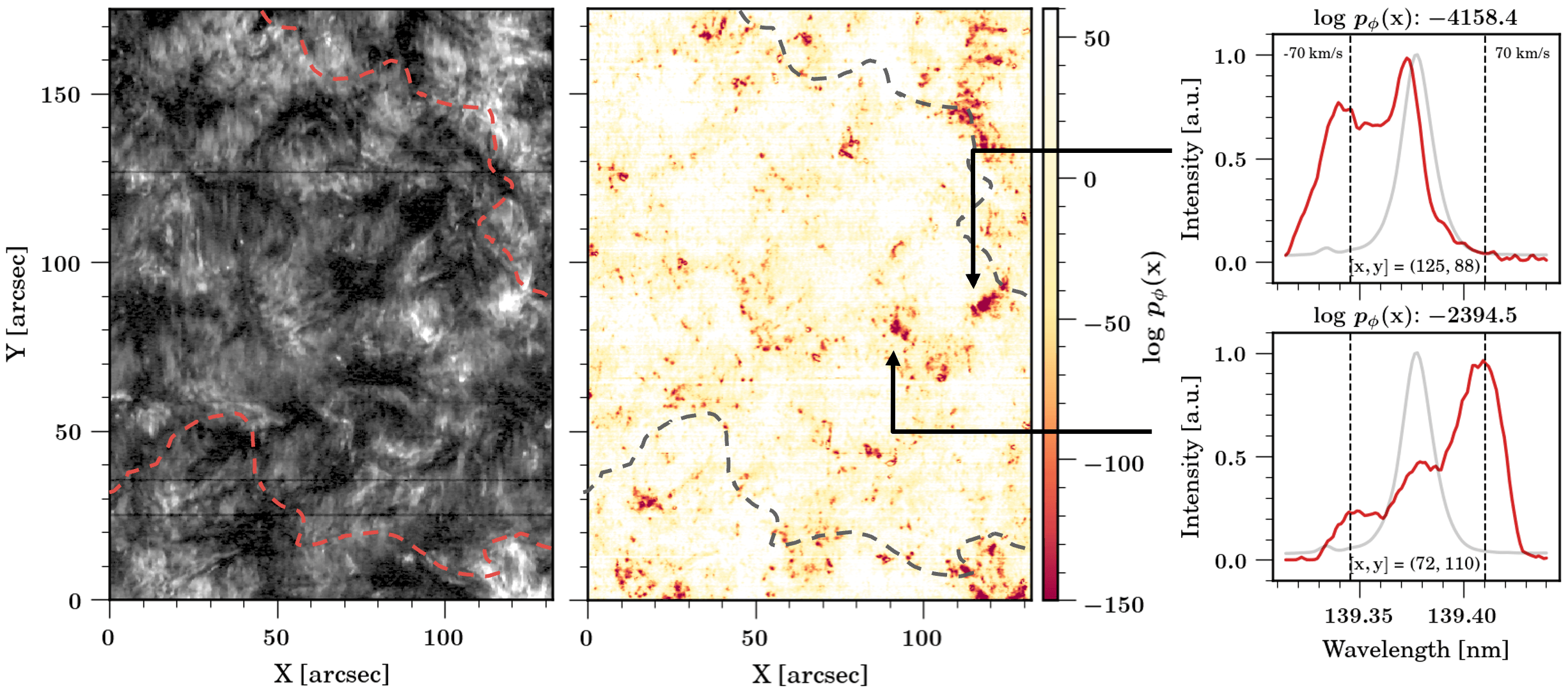}
    \caption{Application of \Inspectorch\ to an IRIS dataset of a coronal hole. The left panels show the intensity map at the core of the \ion{Si}{iv} line and log-probability $\log p_\phi(\xbf)$ map. \fix{Dashed lines indicate the boundaries of the coronal hole as seen in SDO/AIA 193~\AA.} The color scale in the log-probability map is clipped to a minimum of $-150$ to enhance contrast, though some pixels exhibit lower values. The right panel shows an example of two of the most unusual spectra extracted from the low-probability regions. \fix{All spectral profiles are normalized to their own maximum value to facilitate shape comparison (a.u.: arbitrary units); the likelihood evaluation is performed on the unnormalized spectra.} Vertical dashed lines indicate Doppler shifts of $\pm70$~\kms and an average quiet-Sun profile in gray is also shown for reference.}
    \label{fig:iris_logprob}
\end{figure*}

\section{Results: Applications to solar data}\label{sec:results}

The following case studies validate the Normalizing Flow approach across different instruments and physical regimes. All the following results were obtained using \Inspectorch\ to optimize and evaluate the flow models, demonstrating that robust anomaly detection can be achieved with minimal tuning. For each dataset, a separate model is optimized and used to isolate events with very low probability. Implementation details are provided in Appendix~\ref{sec:appendix_implementation}.

\subsection{Detecting supersonic downflows with Hinode/SP}\label{sec:hinode_case}

As a first example, we applied \Inspectorch\ to a Hinode/SP dataset that contains a complex sunspot with multiple light-bridges. Sunspots are complex structures with strong magnetic fields, and identifying rare spectral signatures within them can provide insights into their formation,  \fix{development, and decay processes \citep[e.g.,][]{2003A&ARv..11..153S,2011LRSP....8....4B}}.

This dataset shows the largest sunspot of the active region (AR) NOAA~10953, scanned on April 30, 2007, between 18:35~UT and 19:39~UT, with the spectropolarimeter on the Solar Optical Telescope \citep[SOT/SP;][]{Tsuneta2008,Lites2013} aboard the Hinode satellite  \citep{Kosugi2007}. The data acquisition was performed by sampling the \ion{Fe}{i} 630~nm line pair between 630.089 and 630.327~nm at steps of 21.4~m\AA\ using the normal map mode. The time duration per slit position was 4.8\,s. The FOV is $\sim164^{\prime\prime}\times120^{\prime\prime}$ at $\sim0.16\arcsec$\,pixel$^{-1}$, centered at $(-190^{\prime\prime}, -90^{\prime\prime})$. 

For the sake of simplicity in this first example, we restricted our analysis to the intensity of a single spectral line \fix{\ion{Fe}{i} 630.15~nm, sampling the spectral range between 630.089 and 630.220~nm in $N_{\lambda} = 62$ spectral points}. We consider all pixels to be independent, so the dataset consists of a matrix with size ($N_x \times N_y$, $N_\lambda$). Once the Normalizing Flow is trained, we evaluate it on all pixels in the FOV to construct a probability map. Figure~\ref{fig:hinode_logprob} shows this likelihood map, highlighting the profiles with very low probabilities in red. These regions are extremely scarce, representing only $\sim0.14\%$ of all pixels (defined here as $\log p_\phi(\xbf) < -100$).

By analyzing the spatial distribution of those pixels in Fig.~\ref{fig:hinode_logprob}, we find they mainly appear at the boundary between the sunspot penumbra and the surrounding granulation. These profiles correspond to regions with two components: a superposition of two velocity components of the spectral line, Doppler-shifted relative to each other. One of these components often exhibits very strong downflows of the order of 10~\kms. Two of those profiles are extracted and shown in red in the rightmost panels of Fig.~\ref{fig:hinode_logprob}. These profiles are best interpreted as two-component spectra rather than emission-core profiles, as strong velocity gradients within the resolution element are far more plausible than the extreme heating required to produce photospheric emission \fix{\citep[][]{Borrero2013}}.

This is not the first time that mass flows exceeding the photospheric sound speed ($\sim6$~\kms) have been reported. Indeed, since the launch of Hinode, studies have reported the presence of strong downflows \fix{around 7-14~\kms} in the photosphere \fix{\citep[e.g.,][]{2007ApJ...668L..91B, Shimizu2008}. In particular, the downflows in the periphery of sunspots} were interpreted as representing the formation of a strong field concentration through the process of convective collapse. All of these detections were based on Stokes~$V$ profiles with a signal excess of about 250$-$400~m\AA\ offset from the line core. 

These downflows are not as evident in single-wavelength intensity images, as lower intensities can easily be confused with intergranular lanes or other darker features in the quiet Sun. To address this, \Inspectorch\ uses the full intensity profile of the spectral line to detect these anomalies. For comparison, we also applied \Inspectorch\ to the Stokes $V$ profiles of the same dataset, and the results are shown in Appendix~\ref{sec:appendix_hinode_stokesv}. The probability map obtained from the Stokes $V$ profiles is similar to the one obtained from the intensity profiles, but also includes other regions with unusual physical processes, such as the presence of strong magnetic field gradients in the line of sight that produce complex Stokes $V$ profiles. This shows that the definition of unusual events as observed in different Stokes profiles can differ, directly reflecting the distinct physical mechanisms they trace.

\subsubsection{Comparison with Isolation Forest}

To assess the performance of \Inspectorch\ relative to commonly used anomaly detection methods, we compare it with Isolation Forest \citep{Liu2008,Liu2012}, a widely adopted, non-parametric algorithm. Isolation Forest identifies anomalies based on how easily individual samples can be isolated through random partitioning of the feature space, without explicitly modeling the underlying data distribution. It has been successfully applied in a variety of scientific contexts, including the identification of error sources in gravity-field data from GRACE \citep{Lasser2024} and serves as the default method in other prominent astrophysical frameworks, such as \textit{Astronomaly} \citep{2021A&C....3600481L}.

To explicitly demonstrate how both methods handle spectral correlations, we designed a controlled bottom-up experiment using the Hinode sunspot dataset. In the first stage, both \Inspectorch\ and Isolation Forest were provided with a single intensity value sampled from the line wing. In the second stage, an additional intensity value from the line core was included, introducing correlated spectral information originating from different atmospheric heights. Figure~\ref{fig:hinode_comparison_nflows} shows the log-probability maps produced by \Inspectorch\ (top panels) and the anomaly score maps produced by Isolation Forest (bottom panels). The left column shows the results using a single spectral point, while the right column corresponds to the two-point case.


For \Inspectorch, the inclusion of the second spectral point leads to a clear refinement of the probability map, revealing more coherent and physically meaningful structures. \fixii{For example, the umbra becomes less unusual because it is highly homogeneous and self-similar}. In contrast, the anomaly map produced by Isolation Forest remains largely unchanged when the additional information is included. This behavior reflects a fundamental difference between the two approaches: while Isolation Forest relies on feature-wise partitioning, the Normalizing Flow architecture underlying \Inspectorch\ explicitly models the joint distribution, allowing subtle, physically relevant correlations to influence anomaly detection. Furthermore, when provided with the full spectral profile, the anomaly map produced by Isolation Forest remains virtually identical to the two-point case. This indicates that the algorithm struggles to extract additional meaningful information from the higher-dimensional space, whereas \Inspectorch\ continues to refine its structural probability map as more correlated wavelengths are introduced \citep{2023A&A...673A..35D}.

\subsection{Probing the transition region with IRIS}\label{sec:iris_dataset}

To demonstrate the versatility of \Inspectorch\ across different atmospheric regimes, we applied the method to observations of a coronal hole obtained with the Interface Region Imaging Spectrograph \citep[IRIS;][]{DePontieu2014}. The data consist of a raster scan acquired on October 9, 2013, between 23:26~UT on October 9 and 02:56~UT on October 10, centered at (511\arcsec, 296\arcsec). These observations encompass several ultraviolet spectral lines formed in the chromosphere and transition region, providing access to highly dynamic plasma environments.

In this example, we focus on the \ion{Si}{iv} 139.4~nm line, relevant for studying dynamic events in the solar chromosphere and transition region \citep[e.g.,][]{Cho_2024,2026A&A...706A..70M}. Although only a single line is used here, the method naturally allows for the concatenation of multiple spectral lines to jointly probe different atmospheric heights. As in the previous cases, each pixel is treated as an independent sample, and the optimized Normalizing Flow is evaluated across the full FOV.

Figure~\ref{fig:iris_logprob} shows the resulting log-probability and intensity maps for this IRIS observation. The log-probability map reveals that the most unusual spectral profiles are predominantly located in the magnetic network, i.e., near the boundaries of supergranular cells, where \ion{Si}{iv} emission is typically strongest \fix{due to a variety of highly dynamic transition-region phenomena \citep[e.g., network jets and explosive events;][]{Panesar2018, Chen2019, Faber2022}}. Pixels with a log-probability $\log p_\phi(\xbf)< -150$ represent approximately 2.5~\% of the field of view. Importantly, the low probabilities are not driven solely by high intensities, but by spectra with large Doppler shifts and asymmetric line profiles. These correspond to strong upflows and downflows with velocities of up to $\sim70$~\kms, as illustrated by the two example spectra extracted from the anomalous regions.

The observed 20$-$30~Mm spacing of the bright \ion{Si}{iv} network in the coronal hole, together with the associated strong Doppler excursions, is consistent with the supergranular-scale organization recently identified by Parker Solar Probe as the source of interchange-reconnection jets contributing to the fast solar wind \citep{Bale2023}. This suggests that the unusual spectra identified by \Inspectorch\ may provide an observational link between small-scale transition region dynamics and the large-scale outflows detected in the heliosphere.

\begin{figure}[t]
    \centering
    \includegraphics[width=\columnwidth]{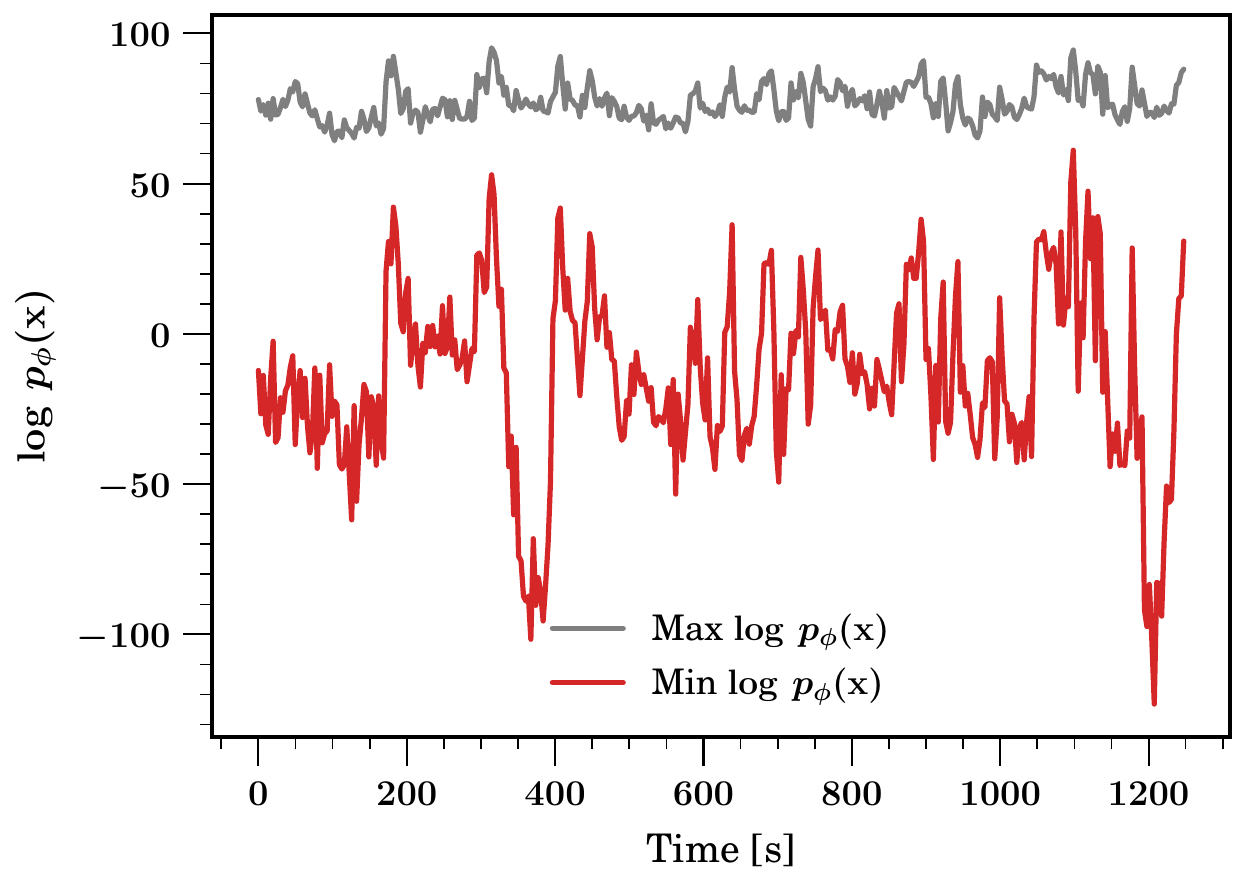}
    \caption{Temporal evolution of the minimum (red line) and maximum (gray line) log-probability values in a time series of MiHI observations. Sharp dips in the minimum log-probability indicate the occurrence of very unusual events.}
    \label{fig:mihi_temporal}
\end{figure}

\begin{figure}[t]
    \centering
    \includegraphics[width=\linewidth]{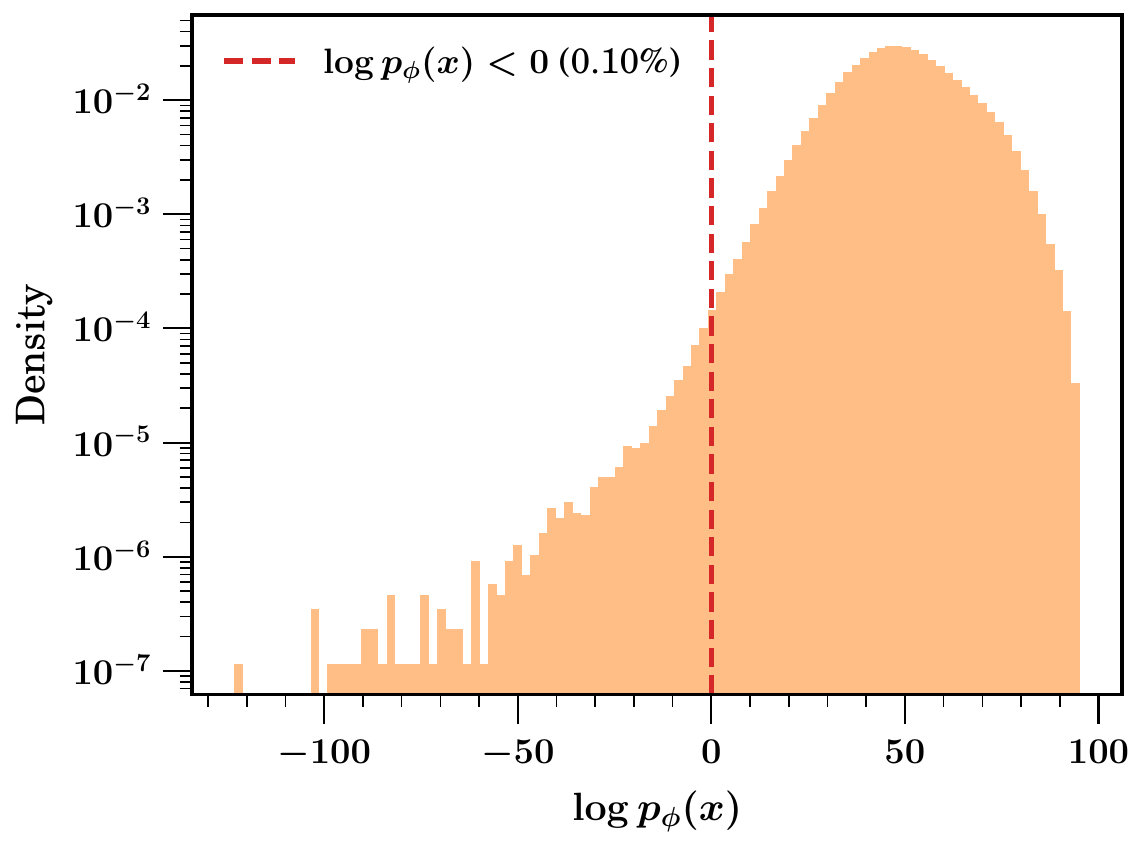}
    \caption{Distribution of the log-probability values in the MiHI dataset. \fix{The percentage in the legend indicates the fraction of all profiles with a log-probability below the vertical line.}}
    \label{fig:mihi_logprob_hist}
\end{figure}

\begin{figure*}[t]
    \sidecaption
    \begin{minipage}{0.99\linewidth}
        \includegraphics[width=\linewidth, trim={0 55px 0 0},clip]{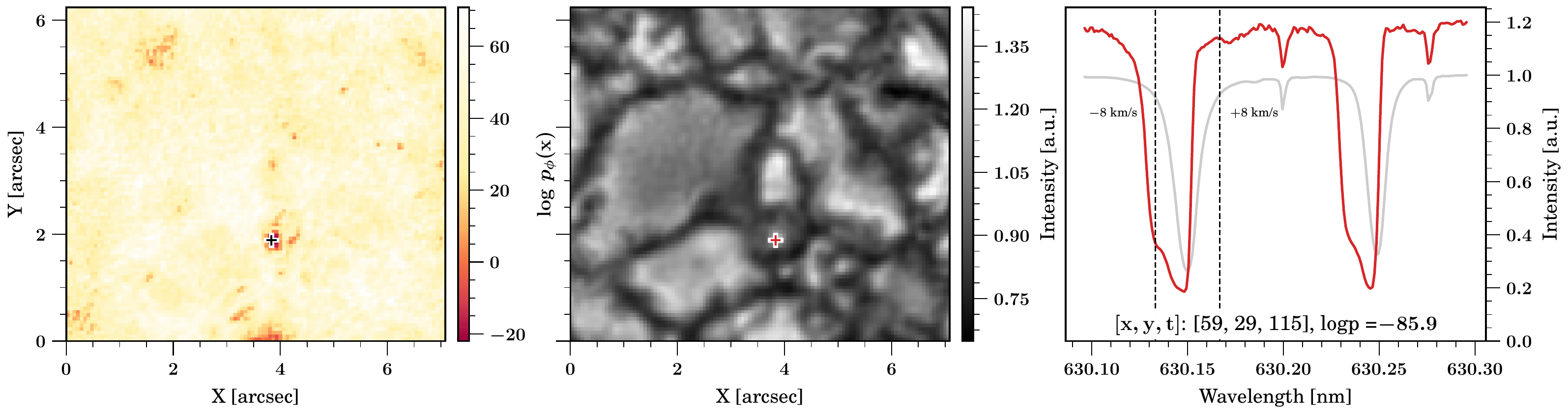}
        \includegraphics[width=\linewidth, trim={0 55px 0 0},clip]{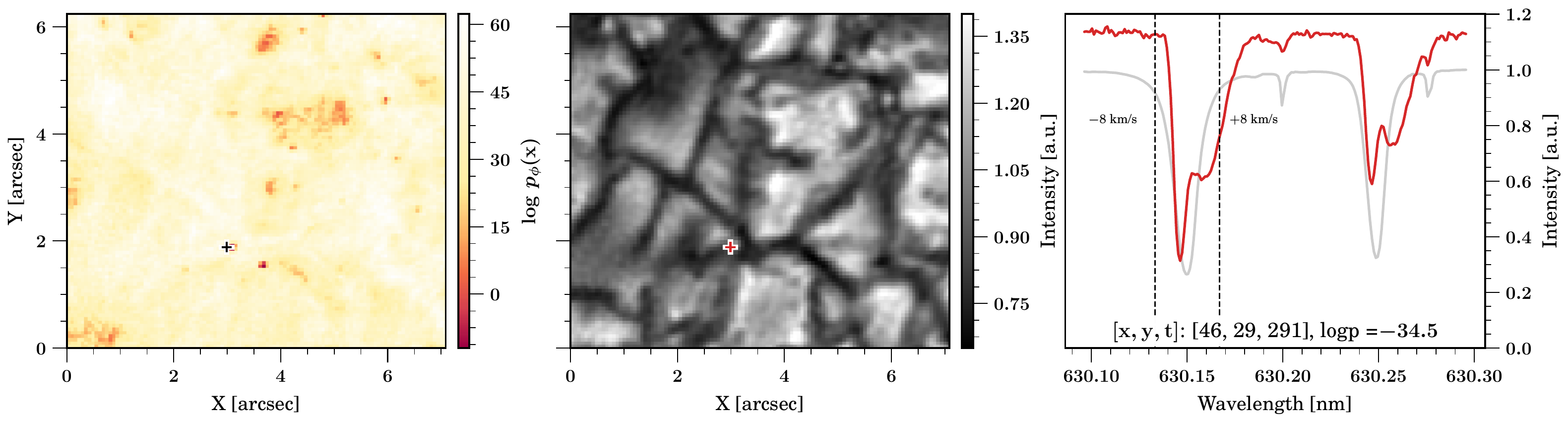}
        \includegraphics[width=\linewidth]{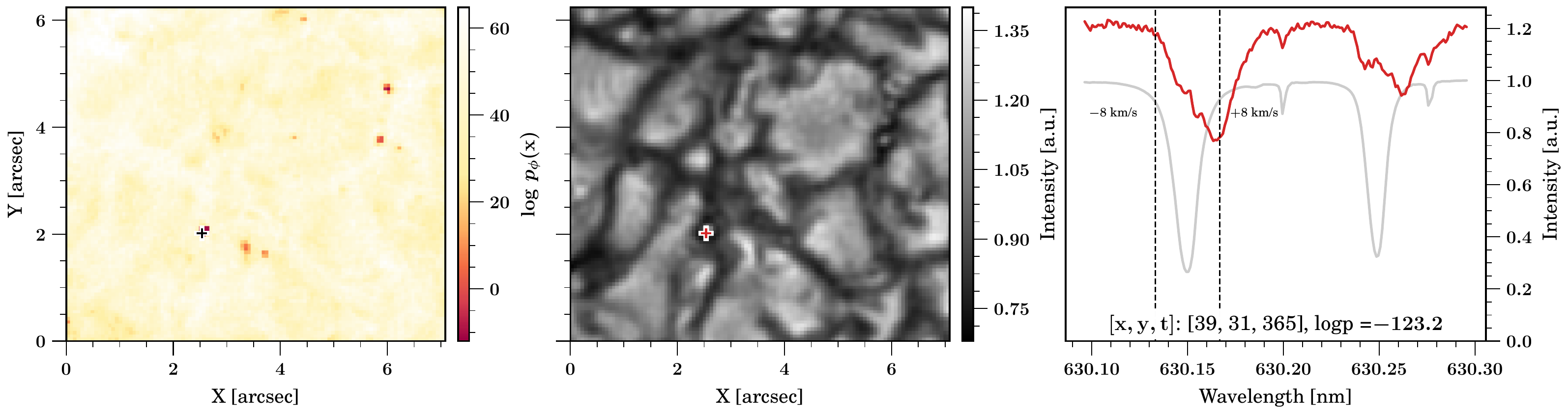}
    \end{minipage}
    \caption{Selection of the three most unusual spectra identified in the MiHI temporal series. Each row shows the log-probability map (left) and the continuum intensity map (middle) at the time of the spectrum (right). A cross marks the location of the spectrum in the maps. The color scale in the log-probability map is clipped to a minimum to enhance contrast. The spectra are normalized to the average quiet Sun intensity at the time of the observation. The gray spectra represent the average quiet Sun profile of this dataset. This figure does not include the full spectral range in MiHI but the inner region around the two \ion{Fe}{i} lines.}
    \label{fig:mihi_3spectra}
\end{figure*}

\subsection{Extreme events in high-cadence MiHI data}

To push our framework to the limits of current observational capabilities, we applied the method to a dataset featuring exceptionally high spatial, spectral, and temporal resolution. The data were recorded with the Microlensed Hyperspectral Imager \citep[MiHI;][]{vanNoort2022A&A...668A.149V}, a prototype integral field spectrograph at the Swedish 1-m Solar Telescope (SST), which provides a full spectrum at every spatial pixel simultaneously. The observations were recorded on August 9, 2018, targeting a quiet-Sun region near disk center. The dataset consists of $128\times114$\,pixels$^2$ with a spatial sampling of 0.065$\arcsec$, 
\fix{covering the \ion{Fe}{i} 630.15 and 630.25~nm\ lines (630.05-630.39~nm) with  
$N_\lambda=340$ spectral points (10~m\AA)}, and a cadence of 3.3\,s over 20.8 minutes.

After training \Inspectorch\ on the full MiHI dataset 
\fix{($N_t, N_x, N_y, N_\lambda$) = (378, 114, 128, 340)},
we obtain a probability estimate for every spectrum at each spatial position and time step. This results in a four-dimensional cube of log-probabilities, enabling a rapid global assessment of extreme events. To identify particularly interesting moments, we compute, for each time step, the minimum log-probability across the field of view. The temporal evolution of these extrema is shown in Fig.~\ref{fig:mihi_temporal}, where the maximum log-probability remains relatively stable, reflecting the dominance of typical spectra, while the minimum log-probability exhibits sharp, isolated dips. These dips indicate the occurrence of highly unusual events confined to specific locations and times.

Applying standard clustering algorithms to an entire dataset often dilutes rare signatures among the vastly more common spectra, requiring an impractical number of clusters to isolate unusual features. Instead, we focus our analysis exclusively on the low-probability tail of the distribution. As shown in the histogram in Fig.~\ref{fig:mihi_logprob_hist}, the log-probabilities exhibit a pronounced tail toward low values. By adopting a threshold of $\log p_\phi(\xbf) < 0$, we isolate the rarest $\sim$0.1\% of all pixels.

\fix{Applying k-means clustering solely to this extreme subset avoids the dilution problem and efficiently reveals three dominant families of unusual spectral profiles.} Figure~\ref{fig:mihi_3spectra} shows a representative example from each of the three families of unusual spectra identified in the MiHI dataset. We have included here the spectral range of the two \ion{Fe}{i} lines, which helps to visualize the different sensitivity of the two lines during the events. All identified spectra represent snapshots of highly dynamic processes rather than isolated events and involve line-of-sight velocities comparable to or exceeding the photospheric sound speed.

The first family (top row) is characterized by very strong upflows reaching velocities of 8~\kms, which are extremely rare in the quiet Sun. These values exceed previously reported photospheric upflows of $\sim4$~\kms\ associated with so-called photospheric jets \citep{MartinezPillet2011, Borrero2013, Jafarzadeh2015}, which have been interpreted in terms of magnetic reconnection between emerging flux and pre-existing fields. While we do not analyze the magnetic properties here, the diverse environments in which these profiles appear suggest that multiple physical mechanisms may be involved.

The second and third families (middle and bottom rows) correspond to strong downflows observed in photospheric bright points. Downflows reaching velocities of 6~\kms\ are commonly associated with the formation and intensification of magnetic elements \citep[e.g.,][]{Nagata2008ApJ, Utz2014ApJ}, consistent with evacuation processes seen in MHD simulations \citep{Danilovic2010}. The third family represents a more extreme and less frequent case, with downflow velocities approaching 8~\kms. In all cases, the detected spectra capture transient phases of evolving processes rather than steady-state phenomena.

This MiHI example highlights the power of likelihood-based anomaly detection, combined with targeted clustering, to efficiently identify and categorize extreme events in high-cadence, high-dimensional solar datasets, paving the way for similar analyses in forthcoming large-scale observations.

\begin{figure*}[t]
    \sidecaption
    \begin{minipage}{\linewidth}
    \includegraphics[width=\linewidth, trim={0 55px 0 0},clip]{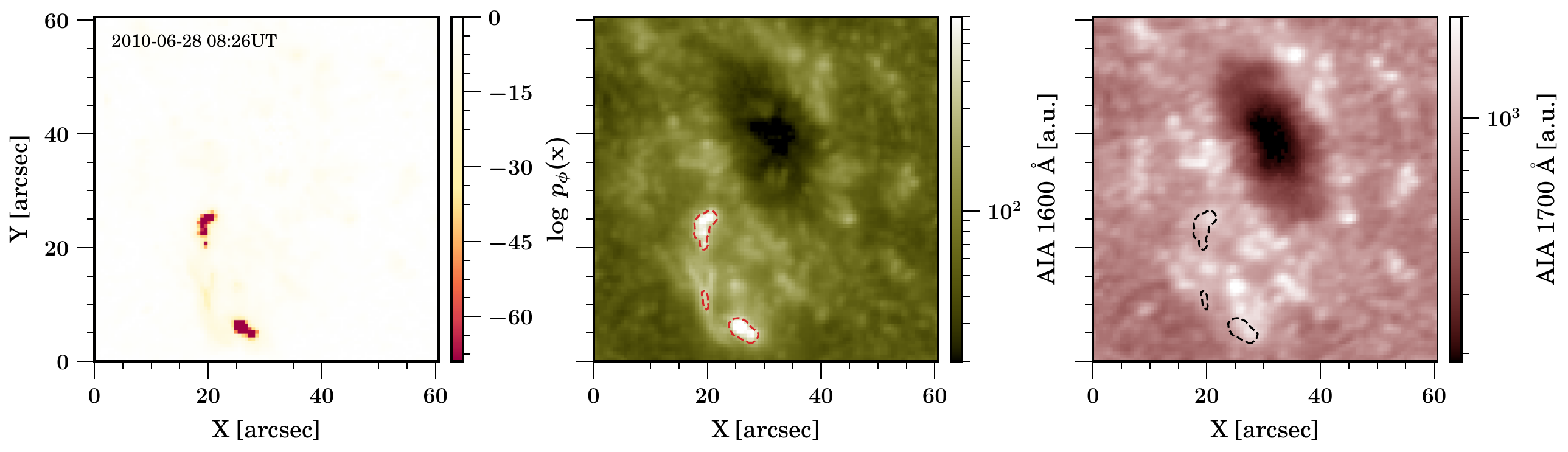}
    \includegraphics[width=\linewidth, trim={0 55px 0 0},clip]{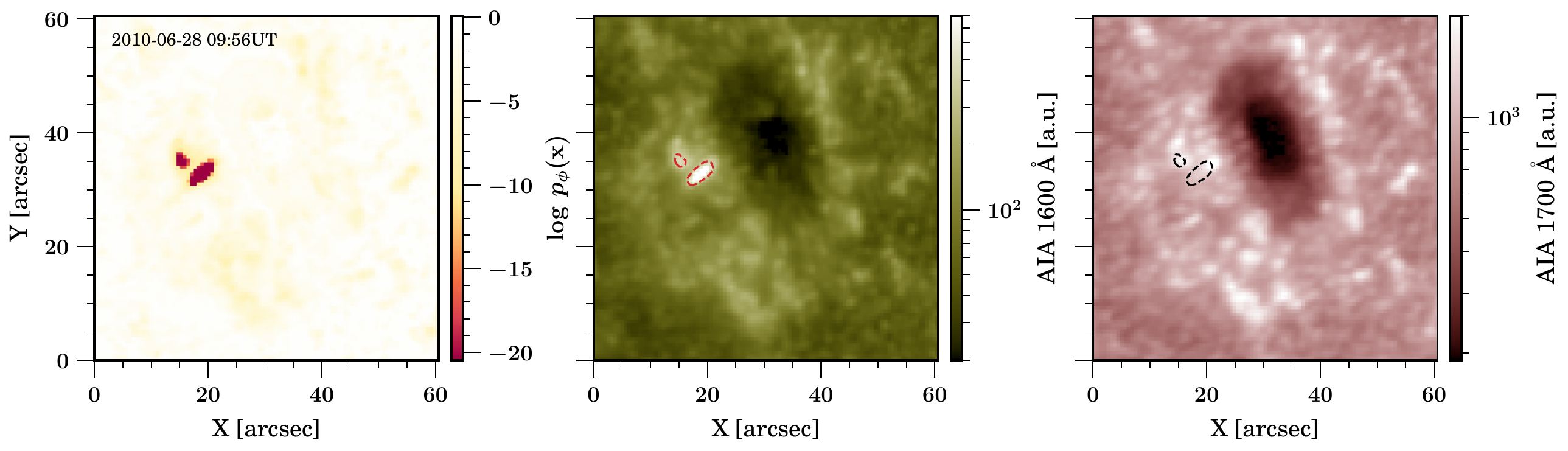}
    \includegraphics[width=\linewidth]{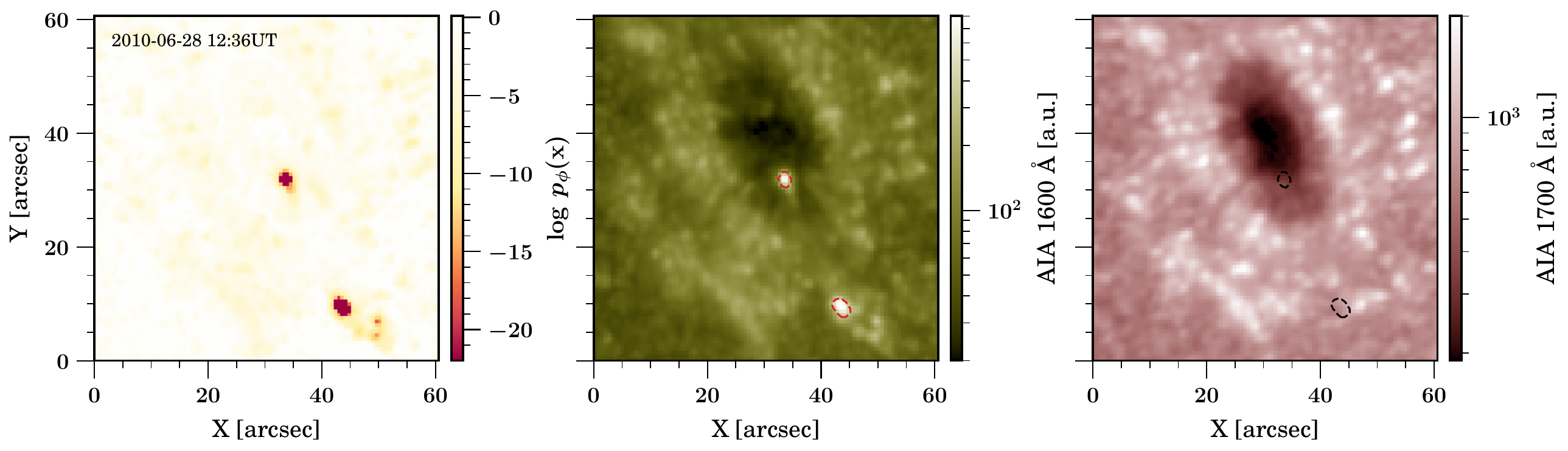}
    \end{minipage}
    \caption{Selection of the three most unusual spectra identified in the SDO dataset when combining the 1600\,\AA\ and 1700\,\AA\ channels. The log-probability map (left) tends to capture the transition-region contribution of the 1600\,\AA\ channel as the unusual behavior. The dashed contours enclose regions where the probability map has a value below $-50,-18,-18$ for each row respectively.}
    \label{fig:sdo_2channels}
\end{figure*}

\subsection{Disentangling multi-channel information with SDO/AIA}
\label{sec:multi_sdo}

A fundamental challenge in analyzing broadband solar imagery is disentangling mixed emission contributions originating from different atmospheric heights. The Atmospheric Imaging Assembly \citep[AIA;][]{aia2012} on board the Solar Dynamics Observatory \citep[SDO;][]{sdo2012} provides \fix{quasi-uninterrupted} full-disk observations of the Sun in multiple passbands, each sampling overlapping atmospheric layers. As a result, individual AIA channels often contain mixed contributions from different heights, which complicates the  interpretation of observed structures.

A well-known example is the similarity between the 1600~\AA\ and 1700~\AA\ passbands, which primarily sample the upper photosphere \citep{fossum_response_2005, rutten_ellerman_2013}. The 1600~\AA\ channel, however, also includes contributions from the transition region through the \ion{C}{iv} doublet, which manifests as elongated, transient brightenings not present in the 1700~\AA\ images \citep{rutten_ellerman_2013, vissers_ellerman_2015}. Distinguishing these overlapping physical contributions presents an ideal scenario for multi-channel density estimation. 

We address this by modeling the joint probability distribution of the 1600~\AA\ and 1700~\AA\ intensities simultaneously. Specifically, we analyze the active region NOAA~11084, observed on 28 June 2010 and centered at ($-$720\arcsec,$-$345\arcsec), previously studied by \citet{2019A&A...626A...4V,SolerPoquet2025}. For each pixel, we concatenate the intensities from the 1600~\AA\ and 1700~\AA\ passbands. The resulting log-probability maps are shown in the first column of Fig.~\ref{fig:sdo_2channels} for three representative time steps, alongside the corresponding 1600~\AA\ and 1700~\AA\ intensity images. The lowest-probability pixels systematically coincide with bright, elongated structures visible in the 1600~\AA\ passband that lack a corresponding signature in 1700~\AA.

This behavior indicates that \Inspectorch\ successfully isolates \fix{intensity-passband} signatures that break the typical correlation between the two passbands. We interpret these low-probability events as originating in the transition region, where enhanced \ion{C}{IV} emission contributes to the 1600~\AA\ channel but not to 1700~\AA. In contrast, the majority of pixels, dominated by upper-photospheric emission, exhibit highly-correlated intensities in both channels and high probabilities are assigned by the model.

This example demonstrates how modeling joint multi-channel distributions enables \Inspectorch\ to disentangle mixed atmospheric contributions and highlight physically distinct phenomena that are difficult to isolate using single-channel or threshold-based analyses.

\subsection{Generalizing to spatial and temporal features}

While the previous sections focused on detecting unusual spectral profiles, the proposed framework is not restricted to spectral information. In general, \Inspectorch\ can model the probability distribution of any set of features extracted from the data, provided they can be represented as vectors. This flexibility allows the method to be naturally extended to spatial, temporal, or spatio-temporal structures, enabling the detection of anomalous patterns beyond purely spectral signatures.

\subsubsection{Spatial anomaly detection in Hinode/SP}

To incorporate spatial information, we model the distribution of local patches extracted from the data. Instead of relying on specialized neural network architectures tailored to spatial data, such as convolutional networks, we directly treat small spatial neighborhoods as multivariate samples drawn from an unknown distribution. By simply flattening these 2D patches into 1D vectors, \Inspectorch\ captures local spatial context while preserving the generality of our standard Normalizing Flow framework.

\begin{figure}[t]
    \centering
    \includegraphics[width=\linewidth]{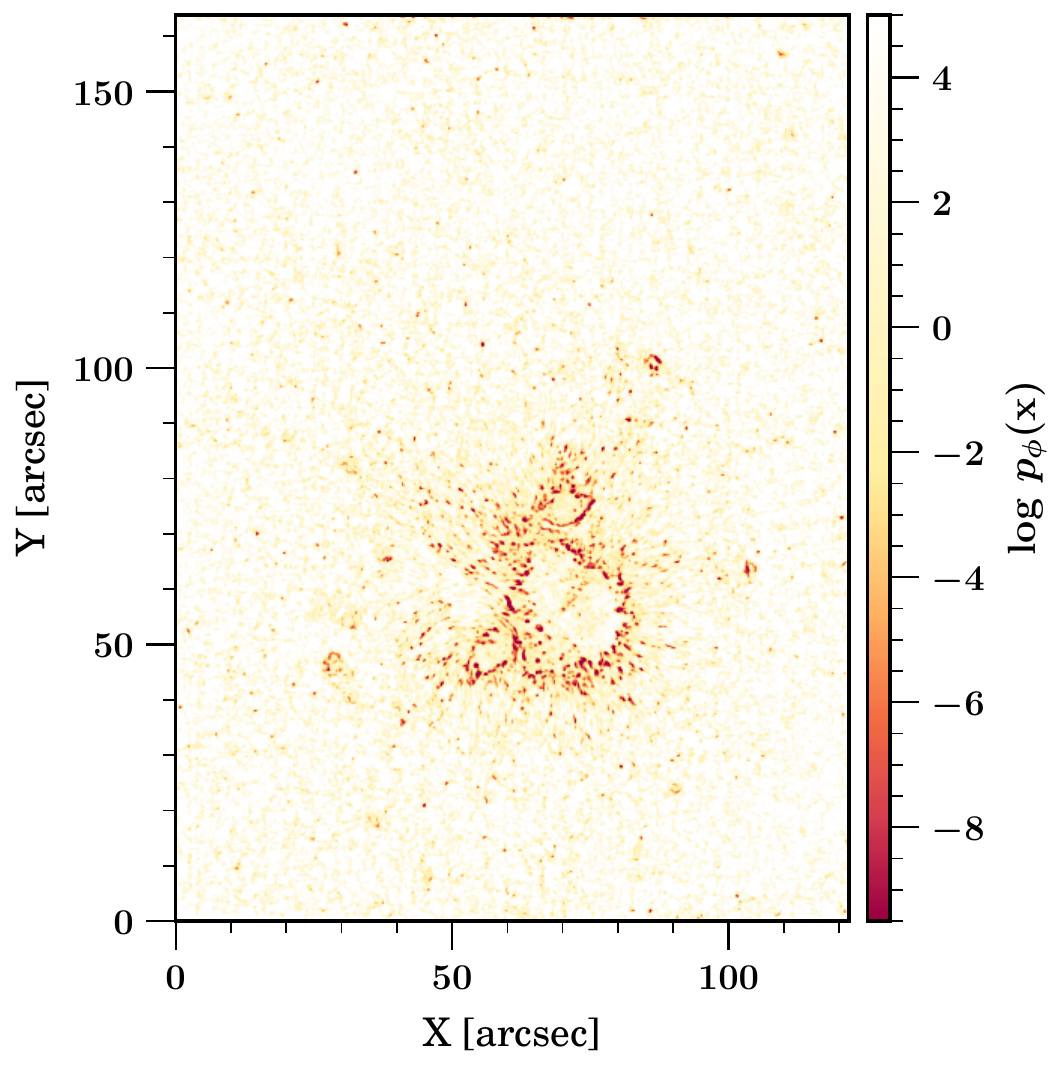}
    \caption{Application of \Inspectorch\ to the spatial dimension of the Hinode/SP dataset, using a \fixii{5 pixels $\times$ 5 pixels patch}. The log-probability map highlights regions with unusual spatial patterns.}
    \label{fig:nflow_spatial}
\end{figure}

An example of this approach is shown in Fig.~\ref{fig:nflow_spatial}, where \Inspectorch\ is applied to the spatial dimension of a Hinode/SP observation. We extract \fixii{5 pixels $\times$ 5 pixels patches} from the data, flatten them into 25-element vectors, and model their distribution using a Normalizing Flow. The resulting log-probability map highlights regions with spatial patterns that deviate from the dominant morphology in the field of view. These regions correspond to localized structures whose spatial organization differs from the more common patterns present in the surrounding atmosphere, providing a data-driven way to identify \fix{unusual spatial features}.

\subsubsection{Temporal anomaly detection}

Beyond spatial information, temporal variability provides an additional and often crucial dimension for distinguishing between different solar phenomena. In many cases, physically distinct processes may appear similar in single snapshots, while exhibiting markedly different temporal evolution. By modeling the distribution of short temporal segments, \Inspectorch\ can identify rare or impulsive dynamical behavior that is not captured by intensity-based or static criteria alone. Similar to the spatial approach, we treat short temporal segments as independent multivariate samples. By flattening these temporal windows into 1D feature vectors, the framework can model their distribution and identify rare or impulsive dynamical behavior that static criteria overlook.

\paragraph{Ellerman bombs vs. network bright points in SDO/AIA:}

A particularly illustrative application of temporal anomaly detection is the separation of Ellerman bombs (EBs) from network bright points using the AIA 1600\,\AA\ and 1700\,\AA\ passbands. As shown by \citet{SolerPoquet2025}, both phenomena can exhibit comparable brightness in these channels, making them difficult to distinguish based on intensity information at a single time step. This ambiguity is further compounded by the fact that both features often occur in strongly magnetized environments.

Despite their similar appearance in individual images, EBs and network bright points exhibit markedly different temporal behavior. EBs are characterized by impulsive light curves with rapid intensity variations and typical lifetimes of approximately 3 minutes \citep{2019A&A...626A...4V}. In contrast, network bright points generally show smoother and more sustained temporal evolution, with typical lifetimes of about one hour \citep{2005ApJ...635..659H}. These differences in temporal dynamics provide a robust physical basis for their separation.

To exploit this distinction, we apply \Inspectorch\ to the temporal evolution of the AIA intensities of the same active region NOAA~11084 of Sect.~\ref{sec:multi_sdo}. The model is trained on temporal windows of 11 consecutive time steps for both the 1600\,\AA\ and 1700\,\AA\ passbands, corresponding to approximately 4.4 minutes of observations. This window length is sufficient to capture the impulsive variability characteristic of EBs while remaining short compared to the typical evolution time of network bright points.

\begin{figure}[t]
    \centering
    \includegraphics[width=0.9\linewidth]{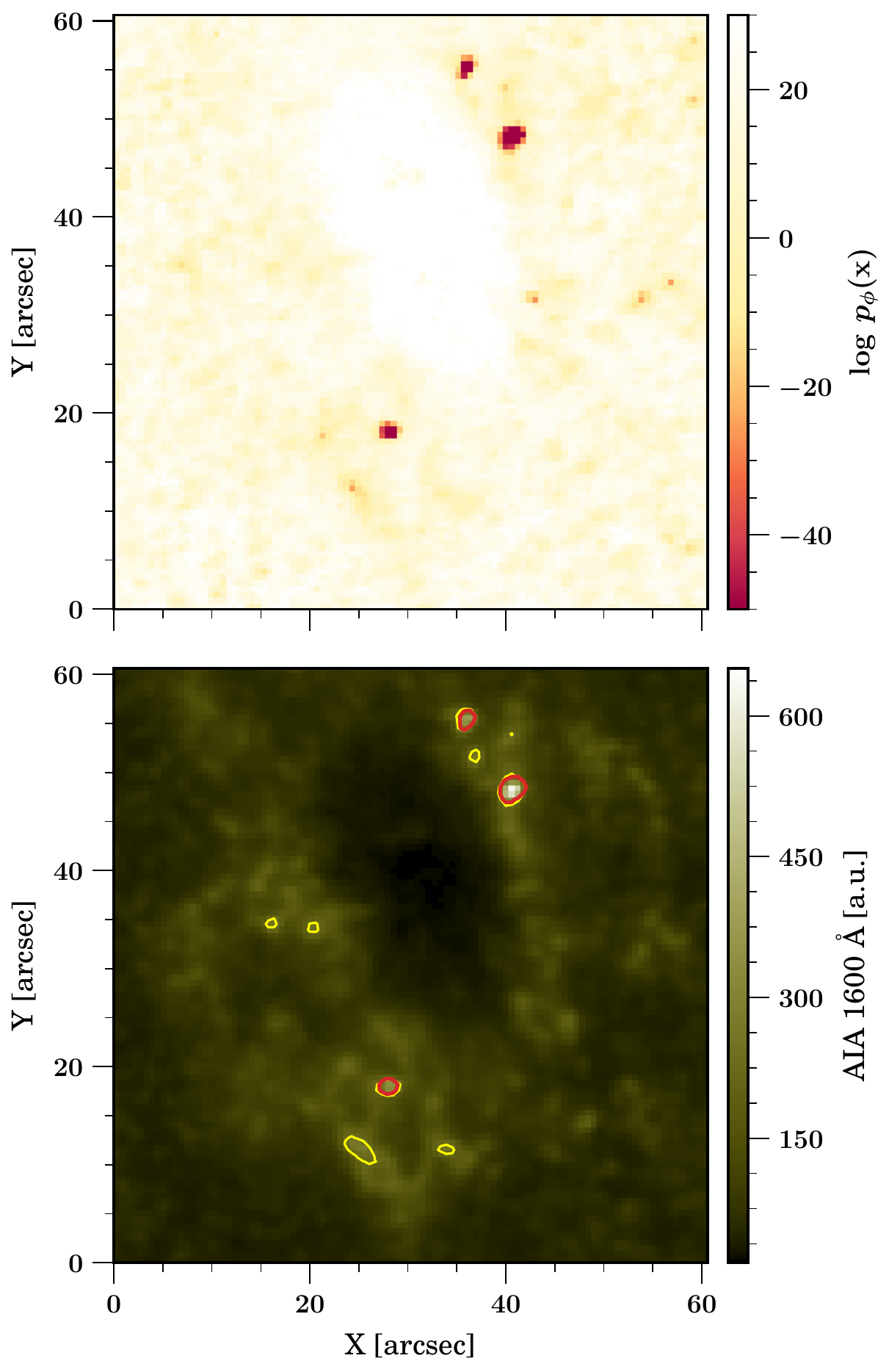}
    \caption{Application of \Inspectorch\ to the temporal dimension of the SDO/AIA dataset using a temporal window of 11 time steps (4.4 minutes). 
    The upper panel shows the log-probability map derived from the temporal evolution of the 1600~\AA\ and 1700~\AA\ passbands, while the lower panel displays the corresponding AIA 1600~\AA\ intensity image at the central frame of the temporal window (approx. 07:50\,UT). The yellow contour marks regions with intensities exceeding three times the average quiet-Sun intensity at the time of the observation. The red contour highlights pixels with log-probability values below $-30$, identifying locations with unusual temporal behavior within the interval. \fix{A movie showing the full temporal evolution is available online.}
}
    \label{fig:sdo_temporal}
\end{figure}

The results of this temporal analysis are shown in Fig.~\ref{fig:sdo_temporal}. This figure \fix{presents} the AIA 1600\,\AA\ around 07:50\,UT and the log-probability map estimated at that moment (with the information of 11 time steps centered at this frame). The log-probability map (upper panel) highlights pixels whose temporal behavior deviates from the dominant patterns in time. These low-probability regions coincide with three isolated events that display impulsive light curves, shown as red contours in the lower panel. For comparison, a simple intensity-based selection (yellow contours) identifies a much larger number of bright features, many of which do not exhibit EB-like temporal behavior. This demonstrates that incorporating temporal information through probabilistic modeling provides a robust, physically grounded framework to distinguish between phenomena that are otherwise identical under static intensity criteria.

\paragraph{Transient coronal events with Solar Orbiter/EUI:}

We further demonstrate the generality of the temporal anomaly detection approach by applying \Inspectorch\ to coronal observations from the Extreme Ultraviolet Imager \citep[EUI;][]{Rochus2020} on board Solar Orbiter \citep{2020A&A...642A...1M}. In the extreme ultraviolet, the corona is dominated by slowly evolving background emission, within which small-scale transient events can be difficult to identify using simple intensity-based criteria.

We have used images from the High Resolution Imager (HRI$_{\mathrm{EUV}}$) at 174\,\AA\ taken on October 9, 2024, between 09:15 and 09:48\,UT with a cadence of 6 seconds. The data were processed to Level~3, including denoising, coalignment, and reprojection, ensuring that the observed temporal variability reflects physical evolution rather than instrumental effects. More details on the data processing can be found in  \citealt{Poirier2025}.

The high cadence of modern solar instruments like EUI can lead to temporal vectors with dimensionalities that are impractical for efficient density estimation. To address this, we employ a dimensionality reduction strategy based on the Fourier transform of the time series. Instead of modeling the raw temporal frames, we calculate the {Power Spectral Density (PSD)} for each pixel. We retain frequencies corresponding to physical timescales ranging from approximately {12 seconds to 11 minutes}. Although this decomposition discards phase information—meaning it identifies anomalous spatial locations without localizing their exact timestamps—it successfully preserves the critical amplitude information required to detect dynamic anomalies while significantly reducing the computational cost and improving the stability of the optimization process. We applied \Inspectorch\ to the PSD vectors and confirmed that the results are consistent with those obtained from the raw time series, but were obtained significantly faster.

\begin{figure}[t]
    \centering
    \includegraphics[width=\linewidth]{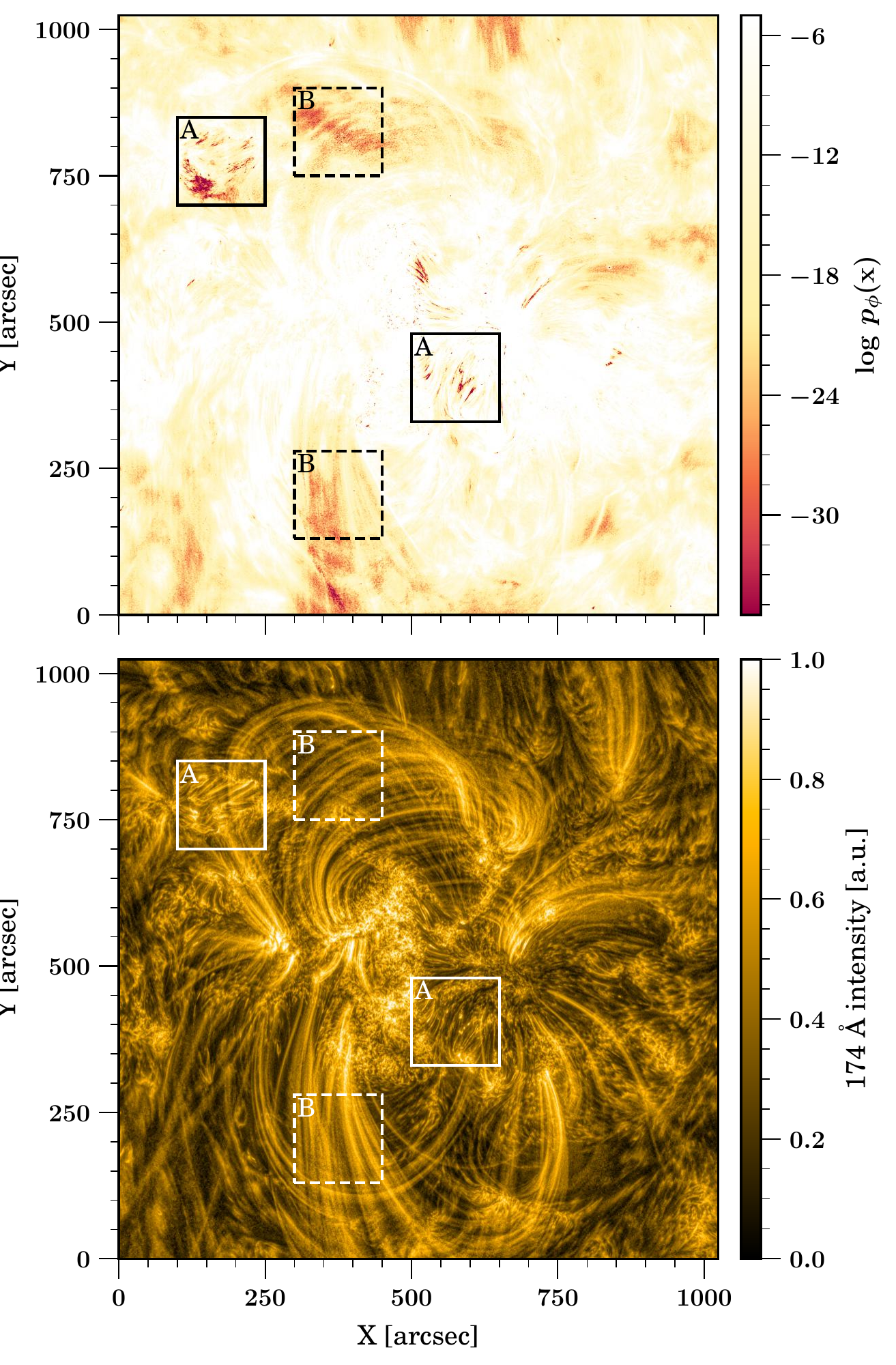}
    \caption{Application of \Inspectorch\ to the temporal dimension of the Solar Orbiter/EUI dataset, using a Fourier decomposition for dimensionality reduction. The upper panel displays the log-probability map of the time series and the lower panel the 174~\AA\ passband at the start of the observations. Solid boxes labeled A highlight regions corresponding to small-scale transient brightenings, while dashed boxes labeled B indicate areas exhibiting coronal loop oscillations. 
    \fix{A movie showing the full temporal evolution is available online.}}
    \label{fig:eui_temporal}
\end{figure}

Figure~\ref{fig:eui_temporal} shows the resulting log-probability map (top panel) for the whole time-series together with an intensity image at the beginning of the observations (bottom panel). The probability map highlights regions with unusual temporal variability, which in this wavelength are often associated with distinct physical processes. Specifically, \Inspectorch\ isolates small-scale transient brightenings \citep{Berghmans2021}, highlighted by the solid boxes labeled A in Fig.~\ref{fig:eui_temporal}, as well as oscillations of coronal loops at larger scales \citep{Poirier2025}, indicated by the dashed boxes labeled B in Fig.~\ref{fig:eui_temporal}. These events exhibit rapid intensity fluctuations compared to the surrounding corona and emerge clearly as low-probability regions in the output, illustrating the ability of the method to isolate diverse dynamic coronal events without relying on ad hoc thresholds or feature-specific tuning.

\section{Scaling density estimation via Flow Matching}\label{sec:flow_matching}

While standard Normalizing Flows effectively solve the anomaly detection problem as demonstrated in previous sections, they impose strict architectural constraints. To guarantee invertibility and efficient Jacobian determinant evaluation (Eq.~\ref{eq:chng_vrbl}), the transformation $\fbf$ must be composed of specialized layers (e.g., coupling layers). 
\fix{As the intrinsic dimensionality of the observations grows and overall dataset volumes approach the petabyte scale}
these constraints can limit expressivity or 
\fix{become prohibitively slow to train at scale.}

To address this, we explore Flow Matching \citep{Lipman2022}, a method for training Continuous Normalizing Flows that we have also integrated into the \Inspectorch\ framework. This approach generalizes the discrete sequence of transformations described in Sect.~\ref{sec:nflows} into a continuous-time evolution. By relaxing architectural constraints while preserving the rigorous probabilistic framework, Flow Matching is rapidly gaining traction across various scientific domains \citep{2025arXiv251014202V}.

\subsection{From discrete layers to continuous vector fields}

We generalize the discrete sequence of transformations described in Sect.~\ref{sec:nflows} into a continuous-time evolution. We define a probability density path $p_t(\mathbf{x})$, parameterized by time $t \in [0,1]$. This path interpolates between a simple base distribution at $t=0$ (denoted $p_0(\mathbf{x})$, equivalent to $p_Z(\mathbf{z})$ in Sect.~\ref{sec:nflows}) and the target data distribution at $t=1$ (denoted $p_1(\mathbf{x})$, equivalent to $p_X(\mathbf{x})$).

The evolution of a sample $\mathbf{x}(t)$ along this path is governed by a time-dependent vector field $\mathbf{v}_\phi(\mathbf{x}, t)$, parameterized by a neural network with parameters $\phi$. The trajectory is defined by the ordinary differential equation (ODE):
\begin{equation}
    \frac{d\mathbf{x}(t)}{dt} = \mathbf{v}_\phi(\mathbf{x}(t), t).
\end{equation}
The final data sample is obtained by integrating this flow from $t=0$ to $t=1$:
\begin{equation}
\mathbf{x}(1) = \mathbf{x}(0) + \int_0^1 \mathbf{v}_\phi(\mathbf{x}(t), t) \, dt.
\end{equation}

This formulation is the continuous analogue to the discrete composition. The change in log-probability is given by the continuous change of variables formula \citep{Chen2018}:
\begin{equation}\label{eq:chng_vrbl_continuous}
\log p_1(\mathbf{x}(1)) = \log p_0(\mathbf{x}(0)) - \int_0^1 \mathrm{Tr}\left( \frac{\partial \mathbf{v}_\phi(\mathbf{x}(t), t)}{\partial \mathbf{x}(t)} \right) dt.
\end{equation}

Comparing Eq.~\ref{eq:chng_vrbl_log} and Eq.~\ref{eq:chng_vrbl_continuous} reveals the fundamental trade-off. Continuous flows allow $\vbf_\phi$ to be modeled by any neural network without invertibility constraints, but require numerical ODE integration to compute likelihoods.

Standard Continuous Normalizing Flows were historically slow to train because solving the ODE and computing the integral at every training step is computationally expensive. Flow Matching circumvents this by learning a vector field that constructs a known path between noise and data. The most popular choice is Conditional Flow Matching with optimal transport paths, where the trajectory between a data sample $\mathbf{x}_{1}$ and a noise sample $\mathbf{x}_{0}$ is a straight line. This implies a constant target velocity $\mathbf{u}_t = \mathbf{x}_{1} - \mathbf{x}_{0}$. The training objective simplifies to a least-squares regression. For a batch of $N$ training triplets $\{\mathbf{x}_{1,i}, \mathbf{x}_{0,i}, t_i\}_{i=1}^N$, where $\mathbf{x}_{1,i}$ is drawn from the dataset, $\mathbf{x}_{0,i}$ from the base distribution, and $t_i$ from $[0,1]$, the loss is:

\begin{equation}
    \mathcal{L}_{\mathrm{FM}}(\phi)
    = \frac{1}{N} \sum_{i=1}^{N}
    \left\| \mathbf{v}_\phi(\mathbf{\psi}_i, t_i)
    - (\mathbf{x}_{1,i} - \mathbf{x}_{0,i}) \right\|^2
    ,
\end{equation}
where $\mathbf{\psi}_i = (1 - t_i)\mathbf{x}_{0,i} + t_i \mathbf{x}_{1,i}$ is the location of the sample at time $t_i$. This loss function is highly scalable as it requires only simple forward passes of the network without ODE integration.

\subsection{Performance trade-offs}

To quantify the practical implications of this continuous formulation, we evaluated both approaches on the Hinode/SP dataset ($8\times10^5$ spectra). In terms of optimization scalability, Flow Matching required only 0.8 minutes to train—approximately five times faster than the discrete Normalizing Flow (4.5 minutes)—benefiting from its unconstrained architecture and simulation-free objective. However, during inference, computing the log-probabilities for the full dataset took nearly 5 minutes with Flow Matching due to the numerical integration required in Eq.~\ref{eq:chng_vrbl_continuous}, compared to just 10 seconds for the discrete flow.

In summary, based on these trade-offs, we recommend standard Normalizing Flows (Sect.~\ref{sec:nflows}) for datasets that fit within memory (e.g., Hinode, IRIS), where rapid, interactive anomaly detection is a priority. For future petabyte-scale campaigns where training time is the bottleneck, Flow Matching could provide the necessary scalability, though accelerating the inference process will be essential to make it fully viable for routine analysis.

\section{Summary and conclusions}\label{sec:conclusions_summary}

In this study, we introduced \Inspectorch, an open-source tool based on flow-based models—a highly flexible class of probabilistic density estimators designed to efficiently identify rare and unusual events in complex solar observations. The method provides a robust and general framework for anomaly detection by assigning physically meaningful probabilities to individual data samples. We demonstrated its applicability across a broad range of instruments and observing modalities, including Hinode/SP, IRIS, MiHI/SST, SDO/AIA, and EUI/Solar Orbiter. The implementation of \Inspectorch\ is publicly available at \url{https://github.com/cdiazbas/inspectorch}.

Many physically important solar phenomena remain poorly explored because they are rare and difficult to identify within large datasets. The probabilistic framework provided by \Inspectorch\ enables a systematic and unbiased search for such events. Our main findings can be summarized as follows:

\begin{enumerate}
    \item \Inspectorch\ successfully models the full multi-dimensional probability distribution of solar spectra and related observables, allowing each sample to be assigned a quantitative probability. This enables a natural and objective ranking of events by their degree of rarity.
    
    \item Across all datasets examined, the method consistently isolates extreme events occupying only a very small fraction of the data (typically $\sim$0.1\%). These rare samples exhibit distinctive spectral or temporal signatures indicative of extreme physical conditions or unresolved dynamic processes that are often overlooked or diluted by standard analysis techniques such as clustering or threshold-based selection.
    
    \item The probabilistic output of \Inspectorch\ provides actionable guidance for prioritizing scientific follow-up. In particular, it can be used to focus computationally expensive spectropolarimetric inversions \fix{onto targeted sub-regions \citep[e.g.,][]{DiazBaso2021_heating}} on the most unusual and potentially most informative events, enabling physical insight before using significant computational resources. This capability will become increasingly important as dataset sizes continue to grow.
    
    \item A direct comparison with the widely used Isolation Forest algorithm demonstrates that \Inspectorch\ offers a substantially larger dynamic range in anomaly scores through its log-probability estimates. This suggests a higher sensitivity to subtle but physically meaningful deviations from typical behavior.
    
    \item The model architecture requires minimal hyperparameter tuning. All results presented in this work were obtained using the same configuration. However, the choice of input features (including the choice of spectral lines or the size of spatial and temporal patches) remains a physical decision that should be driven by the specific scientific objectives.
    
    \item Finally, our exploratory analysis of Flow Matching highlights a promising path forward for future large-scale applications. While it offers substantial gains in training scalability, inference remains slower due to the required integration. Future work will focus on developing efficient estimators or approximations for the log-likelihood integration, which would allow us to retain the training speed of Flow Matching while recovering the fast inference speeds required for routine solar data analysis.
\end{enumerate}

Flow-based density estimators, as implemented in \Inspectorch, are a powerful addition to the solar physicist's data analysis toolkit. By shifting the focus from predefined feature searches to probabilistic exploration, they enable the discovery of unexpected phenomena in large and complex datasets. As solar observations continue to increase in complexity and dimensionality, probabilistic approaches such as \Inspectorch\ provide a general way to navigate large datasets and systematically identify the most informative and physically extreme events. Since the underlying method is agnostic to the specific physical source, this framework holds significant potential for other astrophysical domains facing similar big data challenges, such as stellar spectroscopy, exoplanet characterization, or large-scale galaxy surveys.

\begin{acknowledgements}

\fix{We would like to thank the anonymous referee for their comments and suggestions.}
CJDB thanks Andres Asensio Ramos and Carolina Cuesta-Lazaro for fruitful discussions at the early stages of this work.
This research was supported by the Research Council of Norway through its Centres of Excellence scheme, project number 262622.
CK acknowledges grants RYC2022-037660-I, funded by MCIN/AEI/10.13039/501100011033 and by "ESF Investing in your future", and PID2024-156066OB-C55, funded by MCIN/AEI/ 10.13039/501100011033 and by “ERDF A way of making Europe”.
The Swedish 1-m Solar Telescope is operated on the island of La Palma by the Institute for Solar Physics of Stockholm University in the Spanish Observatorio del Roque de los Muchachos of the Instituto de Astrofísica de Canarias. The Swedish 1-m Solar Telescope, SST, is co-funded by the Swedish Research Council as a national research infrastructure (registration number 4.3-2021-00169).
We acknowledge the community effort devoted to the development of the following open-source packages that were used in this work: numpy (\url{numpy.org}), matplotlib (\url{matplotlib.org}), scipy (\url{scipy.org}), astropy (\url{astropy.org}) and sunpy (\url{sunpy.org}).
This research has made use of NASA's Astrophysics Data System Bibliographic Services.
\end{acknowledgements}

\bibliographystyle{aa}
\bibliography{references}

@article{Nagata2008ApJ,
	adsurl = {https://ui.adsabs.harvard.edu/abs/2008ApJ...677L.145N},
	author = {{Nagata}, Shin'ichi and {Tsuneta}, Saku and {Suematsu}, Yoshinori and {Ichimoto}, Kiyoshi and {Katsukawa}, Yukio and {Shimizu}, Toshifumi and {Yokoyama}, Takaaki and {Tarbell}, Theodore D. and {Lites}, Bruce W. and {Shine}, Richard A. and {Berger}, Thomas E. and {Title}, Alan M. and {Bellot Rubio}, Luis R. and {Orozco Su{\'a}rez}, David},
	doi = {10.1086/588026},
	journal = {\apjl},
	month = apr,
	number = {2},
	pages = {L145},
	title = {{Formation of Solar Magnetic Flux Tubes with Kilogauss Field Strength Induced by Convective Instability}},
	volume = {677},
	year = 2008}

@article{Utz2014ApJ,
	adsurl = {https://ui.adsabs.harvard.edu/abs/2014ApJ...796...79U},
	archiveprefix = {arXiv},
	author = {{Utz}, D. and {del Toro Iniesta}, J.~C. and {Bellot Rubio}, L.~R. and {Jur{\v{c}}{\'a}k}, J. and {Mart{\'\i}nez Pillet}, V. and {Solanki}, S.~K. and {Schmidt}, W.},
	doi = {10.1088/0004-637X/796/2/79},
	eid = {79},
	eprint = {1411.3240},
	journal = {\apj},
	month = dec,
	number = {2},
	pages = {79},
	primaryclass = {astro-ph.SR},
	title = {{The Formation and Disintegration of Magnetic Bright Points Observed by Sunrise/IMaX}},
	volume = {796},
	year = 2014}

@article{2019A&A...626A...4V,
	adsurl = {https://ui.adsabs.harvard.edu/abs/2019A&A...626A...4V},
	archiveprefix = {arXiv},
	author = {{Vissers}, Gregal J.~M. and {Rouppe van der Voort}, Luc H.~M. and {Rutten}, Robert J.},
	doi = {10.1051/0004-6361/201834811},
	eid = {A4},
	eprint = {1901.07975},
	journal = {\aap},
	month = jun,
	pages = {A4},
	primaryclass = {astro-ph.SR},
	title = {{Automating Ellerman bomb detection in ultraviolet continua}},
	volume = {626},
	year = 2019}

@article{2020A&A...640A..71K,
	adsurl = {https://ui.adsabs.harvard.edu/abs/2020A&A...640A..71K},
	archiveprefix = {arXiv},
	author = {{Kuckein}, C. and {Gonz{\'a}lez Manrique}, S.~J. and {Kleint}, L. and {Asensio Ramos}, A.},
	doi = {10.1051/0004-6361/202038408},
	eid = {A71},
	eprint = {2006.10473},
	journal = {\aap},
	month = aug,
	pages = {A71},
	primaryclass = {astro-ph.SR},
	title = {{Determining the dynamics and magnetic fields in He I 10830 {\r{A}} during a solar filament eruption}},
	volume = {640},
	year = 2020}

@article{2022ApJ...933..145L,
	adsurl = {https://ui.adsabs.harvard.edu/abs/2022ApJ...933..145L},
	archiveprefix = {arXiv},
	author = {{Li}, H. and {del Pino Alem{\'a}n}, T. and {Trujillo Bueno}, J. and {Casini}, R.},
	doi = {10.3847/1538-4357/ac745c},
	eid = {145},
	eprint = {2205.15666},
	journal = {\apj},
	month = jul,
	number = {2},
	pages = {145},
	primaryclass = {astro-ph.SR},
	title = {{TIC: A Stokes Inversion Code for Scattering Polarization with Partial Frequency Redistribution and Arbitrary Magnetic Fields}},
	volume = {933},
	year = 2022}

@article{Bale2023,
	adsurl = {https://ui.adsabs.harvard.edu/abs/2023Natur.618..252B},
	archiveprefix = {arXiv},
	author = {{Bale}, S.~D. and {Drake}, J.~F. and {McManus}, M.~D. and {Desai}, M.~I. and {Badman}, S.~T. and {Larson}, D.~E. and {Swisdak}, M. and {Horbury}, T.~S. and {Raouafi}, N.~E. and {Phan}, T. and {Velli}, M. and {McComas}, D.~J. and {Cohen}, C.~M.~S. and {Mitchell}, D. and {Panasenco}, O. and {Kasper}, J.~C.},
	doi = {10.1038/s41586-023-05955-3},
	eprint = {2208.07932},
	journal = {\nat},
	month = jun,
	number = {7964},
	pages = {252-256},
	primaryclass = {astro-ph.SR},
	title = {{Interchange reconnection as the source of the fast solar wind within coronal holes}},
	volume = {618},
	year = 2023}

@article{Danilovic2010,
	adsurl = {https://ui.adsabs.harvard.edu/abs/2010A&A...509A..76D},
	archiveprefix = {arXiv},
	author = {{Danilovic}, S. and {Sch{\"u}ssler}, M. and {Solanki}, S.~K.},
	doi = {10.1051/0004-6361/200912283},
	eid = {A76},
	eprint = {0910.1211},
	journal = {\aap},
	month = jan,
	pages = {A76},
	primaryclass = {astro-ph.SR},
	title = {{Magnetic field intensification: comparison of 3D MHD simulations with Hinode/SP results}},
	volume = {509},
	year = 2010}

@article{Pang2021,
	adsurl = {https://ui.adsabs.harvard.edu/abs/2020arXiv200702500P},
	archiveprefix = {arXiv},
	author = {{Pang}, Guansong and {Shen}, Chunhua and {Cao}, Longbing and {van den Hengel}, Anton},
	doi = {10.48550/arXiv.2007.02500},
	eid = {arXiv:2007.02500},
	eprint = {2007.02500},
	journal = {arXiv e-prints},
	month = jul,
	pages = {arXiv:2007.02500},
	primaryclass = {cs.LG},
	title = {{Deep Learning for Anomaly Detection: A Review}},
	year = 2020}

@article{Chandola2009,
	address = {New York, NY, USA},
	articleno = {15},
	author = {Chandola, Varun and Banerjee, Arindam and Kumar, Vipin},
	doi = {10.1145/1541880.1541882},
	issn = {0360-0300},
	issue_date = {July 2009},
	journal = {ACM Comput. Surv.},
	month = jul,
	number = {3},
	numpages = {58},
	publisher = {Association for Computing Machinery},
	title = {Anomaly detection: A survey},
	url = {https://doi.org/10.1145/1541880.1541882},
	volume = {41},
	year = {2009}}

@article{2022FrASS...937863W,
	adsurl = {https://ui.adsabs.harvard.edu/abs/2022FrASS...937863W},
	author = {{Wang}, Jingjing and {Luo}, Bingxian and {Liu}, Siqing},
	doi = {10.3389/fspas.2022.1037863},
	eid = {1037863},
	journal = {Frontiers in Astronomy and Space Sciences},
	month = oct,
	pages = {1037863},
	title = {{Precursor identification for strong flares based on anomaly detection algorithm}},
	volume = {9},
	year = 2022}

@article{2023MNRAS.526.3072B,
	adsurl = {https://ui.adsabs.harvard.edu/abs/2023MNRAS.526.3072B},
	archiveprefix = {arXiv},
	author = {{B{\"o}hm}, Vanessa and {Kim}, Alex G. and {Juneau}, St{\'e}phanie},
	doi = {10.1093/mnras/stad2773},
	eprint = {2308.00752},
	journal = {\mnras},
	month = dec,
	number = {2},
	pages = {3072-3087},
	primaryclass = {astro-ph.IM},
	title = {{Fast and efficient identification of anomalous galaxy spectra with neural density estimation}},
	volume = {526},
	year = 2023}

@article{2026A&A...706A..70M,
	adsurl = {https://ui.adsabs.harvard.edu/abs/2026A&A...706A..70M},
	archiveprefix = {arXiv},
	author = {{Murabito}, M. and {Andretta}, V. and {Parenti}, S. and {Kuckein}, C. and {Gonz{\'a}lez Manrique}, S.~J. and {Lezzi}, S.~M. and {Guglielmino}, S.~L.},
	doi = {10.1051/0004-6361/202453463},
	eid = {A70},
	eprint = {2511.01494},
	journal = {\aap},
	month = feb,
	pages = {A70},
	primaryclass = {astro-ph.SR},
	title = {{Velocity field of an active region filament from GRIS infrared He I and IRIS ultraviolet observations}},
	volume = {706},
	year = 2026}

@article{2020A&A...642A...1M,
	adsurl = {https://ui.adsabs.harvard.edu/abs/2020A&A...642A...1M},
	archiveprefix = {arXiv},
	author = {{M{\"u}ller}, D. and {St. Cyr}, O.~C. and {Zouganelis}, I. and {Gilbert}, H.~R. and {Marsden}, R. and {Nieves-Chinchilla}, T. and {Antonucci}, E. and {Auch{\`e}re}, F. and {Berghmans}, D. and {Horbury}, T.~S. and {Howard}, R.~A. and {Krucker}, S. and {Maksimovic}, M. and {Owen}, C.~J. and {Rochus}, P. and {Rodriguez-Pacheco}, J. and {Romoli}, M. and {Solanki}, S.~K. and {Bruno}, R. and {Carlsson}, M. and {Fludra}, A. and {Harra}, L. and {Hassler}, D.~M. and {Livi}, S. and {Louarn}, P. and {Peter}, H. and {Sch{\"u}hle}, U. and {Teriaca}, L. and {del Toro Iniesta}, J.~C. and {Wimmer-Schweingruber}, R.~F. and {Marsch}, E. and {Velli}, M. and {De Groof}, A. and {Walsh}, A. and {Williams}, D.},
	doi = {10.1051/0004-6361/202038467},
	eid = {A1},
	eprint = {2009.00861},
	journal = {\aap},
	month = oct,
	pages = {A1},
	primaryclass = {astro-ph.SR},
	title = {{The Solar Orbiter mission. Science overview}},
	volume = {642},
	year = 2020}

@article{2011SSRv..159...19F,
	adsurl = {https://ui.adsabs.harvard.edu/abs/2011SSRv..159...19F},
	archiveprefix = {arXiv},
	author = {{Fletcher}, L. and {Dennis}, B.~R. and {Hudson}, H.~S. and {Krucker}, S. and {Phillips}, K. and {Veronig}, A. and {Battaglia}, M. and {Bone}, L. and {Caspi}, A. and {Chen}, Q. and {Gallagher}, P. and {Grigis}, P.~T. and {Ji}, H. and {Liu}, W. and {Milligan}, R.~O. and {Temmer}, M.},
	doi = {10.1007/s11214-010-9701-8},
	eprint = {1109.5932},
	journal = {\ssr},
	month = sep,
	number = {1-4},
	pages = {19-106},
	primaryclass = {astro-ph.SR},
	title = {{An Observational Overview of Solar Flares}},
	volume = {159},
	year = 2011}

@article{2021A&C....3600481L,
	adsurl = {https://ui.adsabs.harvard.edu/abs/2021A&C....3600481L},
	archiveprefix = {arXiv},
	author = {{Lochner}, M. and {Bassett}, B.~A.},
	doi = {10.1016/j.ascom.2021.100481},
	eid = {100481},
	eprint = {2010.11202},
	journal = {Astronomy and Computing},
	month = jul,
	pages = {100481},
	primaryclass = {astro-ph.IM},
	title = {{ASTRONOMALY: Personalised active anomaly detection in astronomical data}},
	volume = {36},
	year = 2021}

@article{SolerPoquet2025,
	adsurl = {https://ui.adsabs.harvard.edu/abs/2025A&A...699A..54S},
	archiveprefix = {arXiv},
	author = {{Soler Poquet}, I.~J. and {D{\'\i}az Baso}, C.~J. and {Rouppe van der Voort}, L.~H.~M. and {Vissers}, G.~J.~M.},
	doi = {10.1051/0004-6361/202453524},
	eid = {A54},
	eprint = {2505.03023},
	journal = {\aap},
	month = jul,
	pages = {A54},
	primaryclass = {astro-ph.SR},
	title = {{Automatic detection of Ellerman bombs using deep learning}},
	volume = {699},
	year = 2025}

@article{Bhatnagar2024,
	adsurl = {https://ui.adsabs.harvard.edu/abs/2024A&A...689A.156B},
	archiveprefix = {arXiv},
	author = {{Bhatnagar}, Aditi and {Rouppe van der Voort}, Luc and {Joshi}, Jayant},
	doi = {10.1051/0004-6361/202450070},
	eid = {A156},
	eprint = {2406.09585},
	journal = {\aap},
	month = sep,
	pages = {A156},
	primaryclass = {astro-ph.SR},
	title = {{Transition region response to quiet-Sun Ellerman bombs}},
	volume = {689},
	year = 2024}

@article{DePontieu2014,
	adsurl = {https://ui.adsabs.harvard.edu/abs/2014SoPh..289.2733D},
	archiveprefix = {arXiv},
	author = {{De Pontieu}, B. and {Title}, A.~M. and {Lemen}, J.~R. and {Kushner}, G.~D. and {Akin}, D.~J. and {Allard}, B. and {Berger}, T. and {Boerner}, P. and {Cheung}, M. and {Chou}, C. and {Drake}, J.~F. and {Duncan}, D.~W. and {Freeland}, S. and {Heyman}, G.~F. and {Hoffman}, C. and {Hurlburt}, N.~E. and {Lindgren}, R.~W. and {Mathur}, D. and {Rehse}, R. and {Sabolish}, D. and {Seguin}, R. and {Schrijver}, C.~J. and {Tarbell}, T.~D. and {W{\"u}lser}, J. -P. and {Wolfson}, C.~J. and {Yanari}, C. and {Mudge}, J. and {Nguyen-Phuc}, N. and {Timmons}, R. and {van Bezooijen}, R. and {Weingrod}, I. and {Brookner}, R. and {Butcher}, G. and {Dougherty}, B. and {Eder}, J. and {Knagenhjelm}, V. and {Larsen}, S. and {Mansir}, D. and {Phan}, L. and {Boyle}, P. and {Cheimets}, P.~N. and {DeLuca}, E.~E. and {Golub}, L. and {Gates}, R. and {Hertz}, E. and {McKillop}, S. and {Park}, S. and {Perry}, T. and {Podgorski}, W.~A. and {Reeves}, K. and {Saar}, S. and {Testa}, P. and {Tian}, H. and {Weber}, M. and {Dunn}, C. and {Eccles}, S. and {Jaeggli}, S.~A. and {Kankelborg}, C.~C. and {Mashburn}, K. and {Pust}, N. and {Springer}, L. and {Carvalho}, R. and {Kleint}, L. and {Marmie}, J. and {Mazmanian}, E. and {Pereira}, T.~M.~D. and {Sawyer}, S. and {Strong}, J. and {Worden}, S.~P. and {Carlsson}, M. and {Hansteen}, V.~H. and {Leenaarts}, J. and {Wiesmann}, M. and {Aloise}, J. and {Chu}, K. -C. and {Bush}, R.~I. and {Scherrer}, P.~H. and {Brekke}, P. and {Martinez-Sykora}, J. and {Lites}, B.~W. and {McIntosh}, S.~W. and {Uitenbroek}, H. and {Okamoto}, T.~J. and {Gummin}, M.~A. and {Auker}, G. and {Jerram}, P. and {Pool}, P. and {Waltham}, N.},
	doi = {10.1007/s11207-014-0485-y},
	eprint = {1401.2491},
	journal = {\solphys},
	month = jul,
	number = {7},
	pages = {2733-2779},
	primaryclass = {astro-ph.SR},
	title = {{The Interface Region Imaging Spectrograph (IRIS)}},
	volume = {289},
	year = 2014}

@article{Borrero2013,
	adsurl = {https://ui.adsabs.harvard.edu/abs/2013ApJ...768...69B},
	archiveprefix = {arXiv},
	author = {{Borrero}, J.~M. and {Mart{\'\i}nez Pillet}, V. and {Schmidt}, W. and {Quintero Noda}, C. and {Bonet}, J.~A. and {del Toro Iniesta}, J.~C. and {Bellot Rubio}, L.~R.},
	doi = {10.1088/0004-637X/768/1/69},
	eid = {69},
	eprint = {1303.2557},
	journal = {\apj},
	month = may,
	number = {1},
	pages = {69},
	primaryclass = {astro-ph.SR},
	title = {{Is Magnetic Reconnection the Cause of Supersonic Upflows in Granular Cells?}},
	volume = {768},
	year = 2013}

@article{MartinezPillet2011,
	adsurl = {https://ui.adsabs.harvard.edu/abs/2011A&A...530A.111M},
	archiveprefix = {arXiv},
	author = {{Mart{\'\i}nez Pillet}, V. and {Del Toro Iniesta}, J.~C. and {Quintero Noda}, C.},
	doi = {10.1051/0004-6361/201015941},
	eid = {A111},
	eprint = {1104.5564},
	journal = {\aap},
	month = jun,
	pages = {A111},
	primaryclass = {astro-ph.SR},
	title = {{Ubiquitous quiet-Sun jets}},
	volume = {530},
	year = 2011}

@article{Shimizu2008,
	adsurl = {https://ui.adsabs.harvard.edu/abs/2008ApJ...680.1467S},
	archiveprefix = {arXiv},
	author = {{Shimizu}, T. and {Lites}, B.~W. and {Katsukawa}, Y. and {Ichimoto}, K. and {Suematsu}, Y. and {Tsuneta}, S. and {Nagata}, S. and {Kubo}, M. and {Shine}, R.~A. and {Tarbell}, T.~D.},
	doi = {10.1086/588775},
	eprint = {0804.1167},
	journal = {\apj},
	month = jun,
	number = {2},
	pages = {1467-1476},
	primaryclass = {astro-ph},
	title = {{Frequent Occurrence of High-Speed Local Mass Downflows on the Solar Surface}},
	volume = {680},
	year = 2008}

@article{Jafarzadeh2015,
	adsurl = {https://ui.adsabs.harvard.edu/abs/2015ApJ...810...54J},
	archiveprefix = {arXiv},
	author = {{Jafarzadeh}, S. and {Rouppe van der Voort}, L. and {de la Cruz Rodr{\'\i}guez}, J.},
	doi = {10.1088/0004-637X/810/1/54},
	eid = {54},
	eprint = {1507.07355},
	journal = {\apj},
	month = sep,
	number = {1},
	pages = {54},
	primaryclass = {astro-ph.SR},
	title = {{Magnetic Upflow Events in the Quiet-Sun Photosphere. I. Observations}},
	volume = {810},
	year = 2015}

@article{Cho_2024,
	author = {Cho, Kyuhyoun and De Pontieu, Bart and Testa, Paola},
	doi = {10.3847/1538-4357/ad7586},
	journal = {The Astrophysical Journal},
	month = {oct},
	number = {1},
	pages = {33},
	publisher = {The American Astronomical Society},
	title = {On the Nature of Nonthermal Broadening of Spectral Lines Observed by IRIS},
	url = {https://dx.doi.org/10.3847/1538-4357/ad7586},
	volume = {975},
	year = {2024}}

@article{vanNoort2022A&A...668A.149V,
	adsurl = {https://ui.adsabs.harvard.edu/abs/2022A&A...668A.149V},
	author = {{van Noort}, M. and {Bischoff}, J. and {Kramer}, A. and {Solanki}, S.~K. and {Kiselman}, D.},
	doi = {10.1051/0004-6361/202243464},
	eid = {A149},
	journal = {\aap},
	month = dec,
	pages = {A149},
	title = {{A prototype of a microlensed hyperspectral imager for solar observations}},
	volume = {668},
	year = 2022}

@article{Kosugi2007,
	adsurl = {https://ui.adsabs.harvard.edu/abs/2007SoPh..243....3K},
	author = {{Kosugi}, T. and {Matsuzaki}, K. and {Sakao}, T. and {Shimizu}, T. and {Sone}, Y. and {Tachikawa}, S. and {Hashimoto}, T. and {Minesugi}, K. and {Ohnishi}, A. and {Yamada}, T. and {Tsuneta}, S. and {Hara}, H. and {Ichimoto}, K. and {Suematsu}, Y. and {Shimojo}, M. and {Watanabe}, T. and {Shimada}, S. and {Davis}, J.~M. and {Hill}, L.~D. and {Owens}, J.~K. and {Title}, A.~M. and {Culhane}, J.~L. and {Harra}, L.~K. and {Doschek}, G.~A. and {Golub}, L.},
	doi = {10.1007/s11207-007-9014-6},
	journal = {\solphys},
	month = jun,
	number = {1},
	pages = {3-17},
	title = {{The Hinode (Solar-B) Mission: An Overview}},
	volume = {243},
	year = 2007}

@article{Lites2013,
	adsurl = {https://ui.adsabs.harvard.edu/abs/2013SoPh..283..579L},
	author = {{Lites}, B.~W. and {Akin}, D.~L. and {Card}, G. and {Cruz}, T. and {Duncan}, D.~W. and {Edwards}, C.~G. and {Elmore}, D.~F. and {Hoffmann}, C. and {Katsukawa}, Y. and {Katz}, N. and {Kubo}, M. and {Ichimoto}, K. and {Shimizu}, T. and {Shine}, R.~A. and {Streander}, K.~V. and {Suematsu}, A. and {Tarbell}, T.~D. and {Title}, A.~M. and {Tsuneta}, S.},
	doi = {10.1007/s11207-012-0206-3},
	journal = {\solphys},
	month = apr,
	number = {2},
	pages = {579-599},
	title = {{The Hinode Spectro-Polarimeter}},
	volume = {283},
	year = 2013}

@article{Tsuneta2008,
	adsurl = {https://ui.adsabs.harvard.edu/abs/2008SoPh..249..167T},
	archiveprefix = {arXiv},
	author = {{Tsuneta}, S. and {Ichimoto}, K. and {Katsukawa}, Y. and {Nagata}, S. and {Otsubo}, M. and {Shimizu}, T. and {Suematsu}, Y. and {Nakagiri}, M. and {Noguchi}, M. and {Tarbell}, T. and {Title}, A. and {Shine}, R. and {Rosenberg}, W. and {Hoffmann}, C. and {Jurcevich}, B. and {Kushner}, G. and {Levay}, M. and {Lites}, B. and {Elmore}, D. and {Matsushita}, T. and {Kawaguchi}, N. and {Saito}, H. and {Mikami}, I. and {Hill}, L.~D. and {Owens}, J.~K.},
	doi = {10.1007/s11207-008-9174-z},
	eprint = {0711.1715},
	journal = {\solphys},
	month = jun,
	number = {2},
	pages = {167-196},
	primaryclass = {astro-ph},
	title = {{The Solar Optical Telescope for the Hinode Mission: An Overview}},
	volume = {249},
	year = 2008}

@article{Lasser2024,
	address = {Cham},
	author = {Lasser, Martin and Zbinden, Jonas and Meyer, Ulrich and Panos, Brandon and Arnold, Daniel and J{\"a}ggi, Adrian},
	booktitle = {Together Again for Geodesy},
	editor = {Freymueller, Jeffrey T. and S{\'a}nchez, Laura},
	isbn = {978-3-031-91167-5},
	pages = {199--206},
	publisher = {Springer Nature Switzerland},
	title = {Automated Anomaly and Outlier Detection in GRACE and GRACE Follow-On Post-Fit Residuals Using Machine Learning},
	url = {http://link.springer.com/chapter/10.1007/1345_2024_283},
	year = {2025}}

@article{Liu2008,
	author = {Liu, Fei Tony and Ting, Kai Ming and Zhou, Zhi-Hua},
	booktitle = {Eighth IEEE International Conference on Data Mining},
	doi = {10.1109/ICDM.2008.17},
	pages = {413-422},
	title = {Isolation Forest},
	url = {https://doi.org/10.1109/ICDM.2008.17},
	year = {2008}}

@article{Liu2012,
	address = {New York, NY, USA},
	articleno = {3},
	author = {Liu, Fei Tony and Ting, Kai Ming and Zhou, Zhi-Hua},
	doi = {10.1145/2133360.2133363},
	issn = {1556-4681},
	issue_date = {March 2012},
	journal = {ACM Trans. Knowl. Discov. Data},
	month = {Mar},
	number = {1},
	numpages = {39},
	publisher = {Association for Computing Machinery},
	title = {Isolation-Based Anomaly Detection},
	url = {https://doi.org/10.1145/2133360.2133363},
	volume = {6},
	year = {2012}}

@article{Ho2020,
	adsurl = {https://ui.adsabs.harvard.edu/abs/2020arXiv200611239H},
	archiveprefix = {arXiv},
	author = {{Ho}, Jonathan and {Jain}, Ajay and {Abbeel}, Pieter},
	doi = {10.48550/arXiv.2006.11239},
	eid = {arXiv:2006.11239},
	eprint = {2006.11239},
	journal = {arXiv e-prints},
	month = jun,
	pages = {arXiv:2006.11239},
	primaryclass = {cs.LG},
	title = {{Denoising Diffusion Probabilistic Models}},
	year = 2020}

@article{Chen2018,
	adsurl = {https://ui.adsabs.harvard.edu/abs/2018arXiv180607366C},
	archiveprefix = {arXiv},
	author = {{Chen}, Ricky T.~Q. and {Rubanova}, Yulia and {Bettencourt}, Jesse and {Duvenaud}, David},
	doi = {10.48550/arXiv.1806.07366},
	eid = {arXiv:1806.07366},
	eprint = {1806.07366},
	journal = {arXiv e-prints},
	month = jun,
	pages = {arXiv:1806.07366},
	primaryclass = {cs.LG},
	title = {{Neural Ordinary Differential Equations}},
	year = 2018}

@article{Lipman2022,
	adsurl = {https://ui.adsabs.harvard.edu/abs/2022arXiv221002747L},
	archiveprefix = {arXiv},
	author = {{Lipman}, Yaron and {Chen}, Ricky T.~Q. and {Ben-Hamu}, Heli and {Nickel}, Maximilian and {Le}, Matt},
	doi = {10.48550/arXiv.2210.02747},
	eid = {arXiv:2210.02747},
	eprint = {2210.02747},
	journal = {arXiv e-prints},
	month = oct,
	pages = {arXiv:2210.02747},
	primaryclass = {cs.LG},
	title = {{Flow Matching for Generative Modeling}},
	year = 2022}

@article{Song2021,
	adsurl = {https://ui.adsabs.harvard.edu/abs/2020arXiv201113456S},
	archiveprefix = {arXiv},
	author = {{Song}, Yang and {Sohl-Dickstein}, Jascha and {Kingma}, Diederik P. and {Kumar}, Abhishek and {Ermon}, Stefano and {Poole}, Ben},
	doi = {10.48550/arXiv.2011.13456},
	eid = {arXiv:2011.13456},
	eprint = {2011.13456},
	journal = {arXiv e-prints},
	month = nov,
	pages = {arXiv:2011.13456},
	primaryclass = {cs.LG},
	title = {{Score-Based Generative Modeling through Stochastic Differential Equations}},
	year = 2020}

@article{Panos2018_flarekmeans,
	adsurl = {https://ui.adsabs.harvard.edu/abs/2018ApJ...861...62P},
	archiveprefix = {arXiv},
	author = {{Panos}, Brandon and {Kleint}, Lucia and {Huwyler}, Cedric and {Krucker}, S{\"a}m and {Melchior}, Martin and {Ullmann}, Denis and {Voloshynovskiy}, Sviatoslav},
	doi = {10.3847/1538-4357/aac779},
	eid = {62},
	eprint = {1805.10494},
	journal = {\apj},
	month = jul,
	number = {1},
	pages = {62},
	primaryclass = {astro-ph.SR},
	title = {{Identifying Typical Mg II Flare Spectra Using Machine Learning}},
	volume = {861},
	year = 2018}

@article{DiazBaso2022_nflows,
	adsurl = {https://ui.adsabs.harvard.edu/abs/2022A&A...659A.165D},
	archiveprefix = {arXiv},
	author = {{D{\'\i}az Baso}, C.~J. and {Asensio Ramos}, A. and {de la Cruz Rodr{\'\i}guez}, J.},
	doi = {10.1051/0004-6361/202142018},
	eid = {A165},
	eprint = {2108.07089},
	journal = {\aap},
	month = mar,
	pages = {A165},
	primaryclass = {astro-ph.SR},
	title = {{Bayesian Stokes inversion with normalizing flows}},
	volume = {659},
	year = 2022}

@inproceedings{Ciuca2022mla,
	adsurl = {https://ui.adsabs.harvard.edu/abs/2022mla..confE..17C},
	archiveprefix = {arXiv},
	author = {{Ciuca}, Ioana and {Ting}, Yuan-Sen},
	booktitle = {Machine Learning for Astrophysics},
	doi = {10.48550/arXiv.2207.02785},
	eid = {17},
	eprint = {2207.02785},
	month = jul,
	pages = {17},
	primaryclass = {astro-ph.SR},
	title = {{Unsupervised Learning for Stellar Spectra with Deep Normalizing Flows}},
	year = 2022}

@article{Kingma2014,
	adsurl = {https://ui.adsabs.harvard.edu/abs/2014arXiv1412.6980K},
	archiveprefix = {arXiv},
	author = {{Kingma}, Diederik P. and {Ba}, Jimmy},
	eid = {arXiv:1412.6980},
	eprint = {1412.6980},
	journal = {arXiv e-prints},
	month = dec,
	pages = {arXiv:1412.6980},
	primaryclass = {cs.LG},
	title = {{Adam: A Method for Stochastic Optimization}},
	year = 2014}

@article{Durkan2019,
	adsurl = {https://ui.adsabs.harvard.edu/abs/2019arXiv190604032D},
	archiveprefix = {arXiv},
	author = {{Durkan}, Conor and {Bekasov}, Artur and {Murray}, Iain and {Papamakarios}, George},
	eid = {arXiv:1906.04032},
	eprint = {1906.04032},
	journal = {arXiv e-prints},
	month = jun,
	pages = {arXiv:1906.04032},
	primaryclass = {stat.ML},
	title = {{Neural Spline Flows}},
	year = 2019}

@article{Muller2018,
	adsurl = {https://ui.adsabs.harvard.edu/abs/2018arXiv180803856M},
	archiveprefix = {arXiv},
	author = {{M{\"u}ller}, Thomas and {McWilliams}, Brian and {Rousselle}, Fabrice and {Gross}, Markus and {Nov{\'a}k}, Jan},
	eid = {arXiv:1808.03856},
	eprint = {1808.03856},
	journal = {arXiv e-prints},
	month = aug,
	pages = {arXiv:1808.03856},
	primaryclass = {cs.LG},
	title = {{Neural Importance Sampling}},
	year = 2018}

@article{Kobyzev2019Review,
	adsurl = {https://ui.adsabs.harvard.edu/abs/2019arXiv190809257K},
	archiveprefix = {arXiv},
	author = {{Kobyzev}, Ivan and {Prince}, Simon J.~D. and {Brubaker}, Marcus A.},
	eid = {arXiv:1908.09257},
	eprint = {1908.09257},
	journal = {arXiv e-prints},
	month = aug,
	pages = {arXiv:1908.09257},
	primaryclass = {stat.ML},
	title = {{Normalizing Flows: An Introduction and Review of Current Methods}},
	year = 2019}

@article{Papamakarios2019_review,
	adsurl = {https://ui.adsabs.harvard.edu/abs/2019arXiv191202762P},
	archiveprefix = {arXiv},
	author = {{Papamakarios}, George and {Nalisnick}, Eric and {Jimenez Rezende}, Danilo and {Mohamed}, Shakir and {Lakshminarayanan}, Balaji},
	eid = {arXiv:1912.02762},
	eprint = {1912.02762},
	journal = {arXiv e-prints},
	month = dec,
	pages = {arXiv:1912.02762},
	primaryclass = {stat.ML},
	title = {{Normalizing Flows for Probabilistic Modeling and Inference}},
	year = 2019}

@article{SainzDalda2019,
	adsurl = {https://ui.adsabs.harvard.edu/abs/2019ApJ...875L..18S},
	archiveprefix = {arXiv},
	author = {{Sainz Dalda}, Alberto and {de la Cruz Rodr{\'\i}guez}, Jaime and {De Pontieu}, Bart and {Go{\v{s}}i{\'c}}, Milan},
	doi = {10.3847/2041-8213/ab15d9},
	eid = {L18},
	eprint = {1904.08390},
	journal = {\apjl},
	month = apr,
	number = {2},
	pages = {L18},
	primaryclass = {astro-ph.SR},
	title = {{Recovering Thermodynamics from Spectral Profiles observed by IRIS: A Machine and Deep Learning Approach}},
	volume = {875},
	year = 2019}

@article{Asensio2019,
	adsurl = {https://ui.adsabs.harvard.edu/abs/2019A&A...626A.102A},
	archiveprefix = {arXiv},
	author = {{Asensio Ramos}, A. and {D{\'\i}az Baso}, C.~J.},
	doi = {10.1051/0004-6361/201935628},
	eid = {A102},
	eprint = {1904.03714},
	journal = {\aap},
	month = jun,
	pages = {A102},
	primaryclass = {astro-ph.SR},
	title = {{Stokes inversion based on convolutional neural networks}},
	volume = {626},
	year = 2019}

@article{Dinh2014,
	adsurl = {https://ui.adsabs.harvard.edu/abs/2014arXiv1410.8516D},
	archiveprefix = {arXiv},
	author = {{Dinh}, Laurent and {Krueger}, David and {Bengio}, Yoshua},
	eid = {arXiv:1410.8516},
	eprint = {1410.8516},
	journal = {arXiv e-prints},
	month = oct,
	pages = {arXiv:1410.8516},
	primaryclass = {cs.LG},
	title = {{NICE: Non-linear Independent Components Estimation}},
	year = 2014}

@misc{nflows,
	author = {Conor Durkan and Artur Bekasov and Iain Murray and George Papamakarios},
	doi = {10.5281/zenodo.4296287},
	month = nov,
	note = {Version v0.14},
	publisher = {Zenodo},
	title = {nflows: normalizing flows in PyTorch},
	url = {https://doi.org/10.5281/zenodo.4296287},
	year = {2020}}

@article{aia2012,
	adsurl = {http://adsabs.harvard.edu/abs/2012SoPh..275...17L},
	author = {{Lemen}, J.~R. and {Title}, A.~M. and {Akin}, D.~J. and {Boerner}, P.~F. and {Chou}, C. and {Drake}, J.~F. and {Duncan}, D.~W. and {Edwards}, C.~G. and {Friedlaender}, F.~M. and {Heyman}, G.~F. and {Hurlburt}, N.~E. and {Katz}, N.~L. and {Kushner}, G.~D. and {Levay}, M. and {Lindgren}, R.~W. and {Mathur}, D.~P. and {McFeaters}, E.~L. and {Mitchell}, S. and {Rehse}, R.~A. and {Schrijver}, C.~J. and {Springer}, L.~A. and {Stern}, R.~A. and {Tarbell}, T.~D. and {Wuelser}, J.-P. and {Wolfson}, C.~J. and {Yanari}, C. and {Bookbinder}, J.~A. and {Cheimets}, P.~N. and {Caldwell}, D. and {Deluca}, E.~E. and {Gates}, R. and {Golub}, L. and {Park}, S. and {Podgorski}, W.~A. and {Bush}, R.~I. and {Scherrer}, P.~H. and {Gummin}, M.~A. and {Smith}, P. and {Auker}, G. and {Jerram}, P. and {Pool}, P. and {Soufli}, R. and {Windt}, D.~L. and {Beardsley}, S. and {Clapp}, M. and {Lang}, J. and {Waltham}, N.},
	doi = {10.1007/s11207-011-9776-8},
	journal = {\solphys},
	month = jan,
	pages = {17--40},
	title = {{The Atmospheric Imaging Assembly (AIA) on the Solar Dynamics Observatory (SDO)}},
	volume = 275,
	year = 2012}

@article{sdo2012,
	adsurl = {http://adsabs.harvard.edu/abs/2012SoPh..275....3P},
	author = {{Pesnell}, W.~D. and {Thompson}, B.~J. and {Chamberlin}, P.~C.},
	doi = {10.1007/s11207-011-9841-3},
	journal = {\solphys},
	month = jan,
	pages = {3--15},
	title = {{The Solar Dynamics Observatory (SDO)}},
	volume = 275,
	year = 2012}

@article{RuizCobo1992,
	adsurl = {http://adsabs.harvard.edu/abs/1992ApJ...398..375R},
	author = {{Ruiz Cobo}, B. and {del Toro Iniesta}, J.~C.},
	doi = {10.1086/171862},
	journal = {\apj},
	month = oct,
	pages = {375--385},
	title = {{Inversion of Stokes profiles}},
	volume = 398,
	year = 1992}

@article{delaCruz2019_STiC,
	adsurl = {https://ui.adsabs.harvard.edu/abs/2019A&A...623A..74D},
	archiveprefix = {arXiv},
	author = {{de la Cruz Rodr{\'\i}guez}, J. and {Leenaarts}, J. and {Danilovic}, S. and {Uitenbroek}, H.},
	doi = {10.1051/0004-6361/201834464},
	eid = {A74},
	eprint = {1810.08441},
	journal = {\aap},
	month = mar,
	pages = {A74},
	primaryclass = {astro-ph.SR},
	title = {{STiC: A multiatom non-LTE PRD inversion code for full-Stokes solar observations}},
	volume = {623},
	year = 2019}

@article{PyTorch,
	adsurl = {https://ui.adsabs.harvard.edu/abs/2019arXiv191201703P},
	archiveprefix = {arXiv},
	author = {{Paszke}, Adam and {Gross}, Sam and {Massa}, Francisco and {Lerer}, Adam and {Bradbury}, James and {Chanan}, Gregory and {Killeen}, Trevor and {Lin}, Zeming and {Gimelshein}, Natalia and {Antiga}, Luca and {Desmaison}, Alban and {K{\"o}pf}, Andreas and {Yang}, Edward and {DeVito}, Zach and {Raison}, Martin and {Tejani}, Alykhan and {Chilamkurthy}, Sasank and {Steiner}, Benoit and {Fang}, Lu and {Bai}, Junjie and {Chintala}, Soumith},
	eid = {arXiv:1912.01703},
	eprint = {1912.01703},
	journal = {arXiv e-prints},
	month = dec,
	pages = {arXiv:1912.01703},
	primaryclass = {cs.LG},
	title = {{PyTorch: An Imperative Style, High-Performance Deep Learning Library}},
	year = 2019}

@article{fossum_response_2005,
	author = {Fossum, Astrid and Carlsson, Mats},
	doi = {10.1086/429614},
	issn = {0004-637X},
	journal = {The Astrophysical Journal},
	month = may,
	note = {ADS Bibcode: 2005ApJ...625..556F},
	pages = {556--562},
	title = {Response {Functions} of the {Ultraviolet} {Filters} of {TRACE} and the {Detectability} of {High}-{Frequency} {Acoustic} {Waves}},
	url = {https://ui.adsabs.harvard.edu/abs/2005ApJ...625..556F},
	urldate = {2024-02-23},
	volume = {625},
	year = {2005}}

@article{rutten_ellerman_2013,
	author = {Rutten, Robert J and Vissers, Gregal J M and Rouppe van der Voort, Luc H M and S{\"u}tterlin, Peter and Vitas, Nikola},
	doi = {10.1088/1742-6596/440/1/012007},
	issn = {1742-6596},
	journal = {Journal of Physics: Conference Series},
	language = {en},
	month = jun,
	pages = {012007},
	shorttitle = {Ellerman bombs},
	title = {Ellerman bombs: fallacies, fads, usage},
	url = {https://iopscience.iop.org/article/10.1088/1742-6596/440/1/012007},
	urldate = {2023-02-07},
	volume = {440},
	year = {2013}}

@article{vissers_ellerman_2015,
	author = {Vissers, G. J. M. and Rouppe van der Voort, L. H. M. and Rutten, R. J. and Carlsson, M. and De Pontieu, B.},
	doi = {10.1088/0004-637X/812/1/11},
	issn = {0004-637X},
	journal = {The Astrophysical Journal},
	month = oct,
	note = {ADS Bibcode: 2015ApJ...812...11V},
	pages = {11},
	title = {Ellerman {Bombs} at {High} {Resolution}. {III}. {Simultaneous} {Observations} with {IRIS} and {SST}},
	url = {https://ui.adsabs.harvard.edu/abs/2015ApJ...812...11V},
	urldate = {2023-04-19},
	volume = {812},
	year = {2015}}

@article{2005ApJ...635..659H,
	adsurl = {https://ui.adsabs.harvard.edu/abs/2005ApJ...635..659H},
	author = {{Hagenaar}, Hermance J. and {Shine}, Richard A.},
	doi = {10.1086/497367},
	journal = {\apj},
	month = dec,
	number = {1},
	pages = {659-669},
	title = {{Moving Magnetic Features around Sunspots}},
	volume = {635},
	year = 2005}

@article{Rochus2020,
	adsurl = {https://ui.adsabs.harvard.edu/abs/2020A&A...642A...8R},
	archiveprefix = {arXiv},
	author = {{Rochus}, P. and {Auch{`e}re}, F. and {Berghmans}, D. and {Harra}, L. and {Schmutz}, W. and {Sch{\"u}hle}, U. and {Addison}, P. and {Appourchaux}, T. and {Aznar Cuadrado}, R. and {Baker}, D. and others},
	doi = {10.1051/0004-6361/201936663},
	eid = {A8},
	eprint = {2009.10696},
	journal = {\aap},
	month = oct,
	pages = {A8},
	primaryclass = {astro-ph.IM},
	title = {{The Solar Orbiter EUI instrument: The Extreme Ultraviolet Imager}},
	volume = {642},
	year = 2020}

@article{Berghmans2021,
	adsurl = {https://ui.adsabs.harvard.edu/abs/2021A&A...656L...4B},
	archiveprefix = {arXiv},
	author = {{Berghmans}, D. and {Auch{`e}re}, F. and {Long}, D.~M. and {Soubri{\'e}}, E. and {Mierla}, M. and {Zhukov}, A.~N. and {Sch{\"u}hle}, U. and {Antolin}, P. and {Harra}, L. and {Parenti}, S. and others},
	doi = {10.1051/0004-6361/202140380},
	eid = {L4},
	eprint = {2104.14489},
	journal = {\aap},
	month = dec,
	pages = {L4},
	primaryclass = {astro-ph.SR},
	title = {{Extreme-UV quiet Sun brightenings observed by the Solar Orbiter/EUI}},
	volume = {656},
	year = 2021}

@article{Poirier2025,
	adsurl = {https://ui.adsabs.harvard.edu/abs/2025A&A...696A.125P},
	archiveprefix = {arXiv},
	author = {{Poirier}, N. and {Danilovic}, S. and {Kohutova}, P. and {D{\'\i}az Baso}, C.~J. and {Rouppe van der Voort}, L. and {Calchetti}, D. and {Sinjan}, J.},
	doi = {10.1051/0004-6361/202453025},
	eid = {A125},
	eprint = {2412.14805},
	journal = {\aap},
	month = apr,
	pages = {A125},
	primaryclass = {astro-ph.SR},
	title = {{Coronal kink oscillations and photospheric driving: Combining SolO/EUI and SST/CRISP high-resolution observations}},
	volume = {696},
	year = 2025}

@article{DiazBaso2021_heating,
	adsurl = {https://ui.adsabs.harvard.edu/abs/2021A&A...647A.188D},
	archiveprefix = {arXiv},
	author = {{D{\'\i}az Baso}, C.~J. and {de la Cruz Rodr{\'\i}guez}, J. and {Leenaarts}, J.},
	doi = {10.1051/0004-6361/202040111},
	eid = {A188},
	eprint = {2012.06229},
	journal = {\aap},
	month = mar,
	pages = {A188},
	primaryclass = {astro-ph.SR},
	title = {{An observationally constrained model of strong magnetic reconnection in the solar chromosphere. Atmospheric stratification and estimates of heating rates}},
	volume = {647},
	year = 2021}

@article{2025arXiv251014202V,
	adsurl = {https://ui.adsabs.harvard.edu/abs/2025arXiv251014202V},
	archiveprefix = {arXiv},
	author = {{Vidal}, Edgar P. and {Gagliano}, Alexander T. and {Cuesta-Lazaro}, Carolina},
	doi = {10.48550/arXiv.2510.14202},
	eid = {arXiv:2510.14202},
	eprint = {2510.14202},
	journal = {arXiv e-prints},
	month = oct,
	pages = {arXiv:2510.14202},
	primaryclass = {astro-ph.IM},
	title = {{Hierarchical Simulation-Based Inference of Supernova Power Sources and their Physical Properties}},
	year = 2025}

@article{2023A&A...673A..35D,
	adsurl = {https://ui.adsabs.harvard.edu/abs/2023A&A...673A..35D},
	archiveprefix = {arXiv},
	author = {{D{\'\i}az Baso}, C.~J. and {Rouppe van der Voort}, L. and {de la Cruz Rodr{\'\i}guez}, J. and {Leenaarts}, J.},
	doi = {10.1051/0004-6361/202346230},
	eid = {A35},
	eprint = {2303.13875},
	journal = {\aap},
	month = may,
	pages = {A35},
	primaryclass = {astro-ph.SR},
	title = {{Designing wavelength sampling for Fabry-P{\'e}rot observations. Information-based spectral sampling}},
	volume = {673},
	year = 2023}

@article{2007ApJ...663.1386P,
	adsurl = {https://ui.adsabs.harvard.edu/abs/2007ApJ...663.1386P},
	archiveprefix = {arXiv},
	author = {{Pietarila}, A. and {Socas-Navarro}, H. and {Bogdan}, T.},
	doi = {10.1086/518714},
	eprint = {0707.1310},
	journal = {\apj},
	month = jul,
	number = {2},
	pages = {1386-1405},
	primaryclass = {astro-ph},
	title = {{Spectropolarimetric Observations of the Ca II {\ensuremath{\lambda}}8498 and {\ensuremath{\lambda}}8542 in the Quiet Sun}},
	volume = {663},
	year = 2007}

@article{2011A&A...530A..14V,
	adsurl = {https://ui.adsabs.harvard.edu/abs/2011A&A...530A..14V},
	archiveprefix = {arXiv},
	author = {{Viticchi{\'e}}, B. and {S{\'a}nchez Almeida}, J.},
	doi = {10.1051/0004-6361/201016096},
	eid = {A14},
	eprint = {1103.1987},
	journal = {\aap},
	month = jun,
	pages = {A14},
	primaryclass = {astro-ph.SR},
	title = {{Asymmetries of the Stokes V profiles observed by HINODE SOT/SP in the quiet Sun}},
	volume = {530},
	year = 2011}

@article{2014LRSP...11....3C,
	adsurl = {https://ui.adsabs.harvard.edu/abs/2014LRSP...11....3C},
	author = {{Cheung}, Mark C.~M. and {Isobe}, Hiroaki},
	doi = {10.12942/lrsp-2014-3},
	eid = {3},
	journal = {Living Reviews in Solar Physics},
	month = dec,
	number = {1},
	pages = {3},
	title = {{Flux Emergence (Theory)}},
	volume = {11},
	year = 2014}

@article{2019LRSP...16....1B,
	adsurl = {https://ui.adsabs.harvard.edu/abs/2019LRSP...16....1B},
	author = {{Bellot Rubio}, Luis and {Orozco Su{\'a}rez}, David},
	doi = {10.1007/s41116-018-0017-1},
	eid = {1},
	journal = {Living Reviews in Solar Physics},
	month = dec,
	number = {1},
	pages = {1},
	title = {{Quiet Sun magnetic fields: an observational view}},
	volume = {16},
	year = 2019}

@article{2011LRSP....8....4B,
	adsurl = {https://ui.adsabs.harvard.edu/abs/2011LRSP....8....4B},
	archiveprefix = {arXiv},
	author = {{Borrero}, Juan M. and {Ichimoto}, Kiyoshi},
	doi = {10.12942/lrsp-2011-4},
	eid = {4},
	eprint = {1109.4412},
	journal = {Living Reviews in Solar Physics},
	month = dec,
	number = {1},
	pages = {4},
	primaryclass = {astro-ph.SR},
	title = {{Magnetic Structure of Sunspots}},
	volume = {8},
	year = 2011}

@article{2003A&ARv..11..153S,
	adsurl = {https://ui.adsabs.harvard.edu/abs/2003A&ARv..11..153S},
	author = {{Solanki}, Sami K.},
	doi = {10.1007/s00159-003-0018-4},
	journal = {\aapr},
	month = jan,
	number = {2-3},
	pages = {153-286},
	title = {{Sunspots: An overview}},
	volume = {11},
	year = 2003}

@article{2007ApJ...668L..91B,
	adsurl = {https://ui.adsabs.harvard.edu/abs/2007ApJ...668L..91B},
	archiveprefix = {arXiv},
	author = {{Bellot Rubio}, L.~R. and {Tsuneta}, S. and {Ichimoto}, K. and {Katsukawa}, Y. and {Lites}, B.~W. and {Nagata}, S. and {Shimizu}, T. and {Shine}, R.~A. and {Suematsu}, Y. and {Tarbell}, T.~D. and {Title}, A.~M. and {del Toro Iniesta}, J.~C.},
	doi = {10.1086/522604},
	eprint = {0708.2791},
	journal = {\apjl},
	month = oct,
	number = {1},
	pages = {L91-L94},
	primaryclass = {astro-ph},
	title = {{Vector Spectropolarimetry of Dark-cored Penumbral Filaments with Hinode}},
	volume = {668},
	year = 2007}

@article{Panesar2018,
	adsurl = {https://ui.adsabs.harvard.edu/abs/2018ApJ...868L..27P},
	archiveprefix = {arXiv},
	author = {{Panesar}, Navdeep K. and {Sterling}, Alphonse C. and {Moore}, Ronald L. and {Tiwari}, Sanjiv K. and {De Pontieu}, Bart and {Norton}, Aimee A.},
	doi = {10.3847/2041-8213/aaef37},
	eid = {L27},
	eprint = {1811.04314},
	journal = {\apjl},
	month = dec,
	number = {2},
	pages = {L27},
	primaryclass = {astro-ph.SR},
	title = {{IRIS and SDO Observations of Solar Jetlets Resulting from Network-edge Flux Cancelation}},
	volume = {868},
	year = 2018}

@article{Chen2019,
	adsurl = {https://ui.adsabs.harvard.edu/abs/2019ApJ...873...79C},
	archiveprefix = {arXiv},
	author = {{Chen}, Yajie and {Tian}, Hui and {Huang}, Zhenghua and {Peter}, Hardi and {Samanta}, Tanmoy},
	doi = {10.3847/1538-4357/ab0417},
	eid = {79},
	eprint = {1901.11215},
	journal = {\apj},
	month = mar,
	number = {1},
	pages = {79},
	primaryclass = {astro-ph.SR},
	title = {{Investigating the Transition Region Explosive Events and Their Relationship to Network Jets}},
	volume = {873},
	year = 2019}

@mastersthesis{Faber2022,
	author = {{Faber}, Jonas Thoen},
	school = {University of Oslo},
	title = {{Characterizing Ultraviolet Bursts in a Solar Coronal Hole using Machine Learning Techniques}},
	year = 2022,
	month = jan,
	url = {https://hdl.handle.net/10852/96748},
	note = {Master thesis},
	address = {Oslo, Norway},
	publisher = {Universitetet i Oslo},
	language = {English}}

\appendix
\section{Implementation details}\label{sec:appendix_implementation}
\Inspectorch\ is a Python library built on \texttt{PyTorch} \citep{PyTorch} that provides a unified interface for both methods. For Normalizing Flows, we rely on the \texttt{nflows} package \citep{nflows}, using Rational-Quadratic Spline coupling layers \citep{Durkan2019} combined with invertible linear transformations (LU-decomposition) and permutations, a strategy that has been shown to be very effective for modeling complex distributions in high-dimensional spaces \citep{DiazBaso2022_nflows}. For Flow Matching, we utilize the \texttt{flow\_matching} library \citep{Lipman2022} with Conditional Optimal Transport paths. The vector field $\vbf_t$ is parameterized by a neural network (e.g., a Residual Neural Network) that takes both the state $\xbf$ and time $t$ as input. Probability densities are computed by integrating the ODE using the \texttt{euler} solver. In both cases, models are trained using the Adam or AdamW optimizer \citep{Kingma2014} with a learning rate of $10^{-4}$--$10^{-3}$ for approximately 10--50 epochs until convergence.

\section{Sequential transformations}\label{sec:appendix_seq_transf}


Among the different families of transformations $\mathbf{f_\phi}$, we use a transformation known as coupling neural splines flows \citep{Dinh2014, Muller2018, Durkan2019} which have been demonstrated to be effective at representing complex densities and are quick to train and quick to evaluate (see \citealt{Papamakarios2019_review} and \citealt{Kobyzev2019Review} for extensive reviews).

The idea behind the coupling transform was introduced by \citet{Dinh2014} and consists of dividing the input variable (of dimension $Q$) into two parts and applying an invertible transformation $\mathbf{g}$ to the second half ($\mathbf{z}_{q+1:Q}$), whose parameters are a function of the first half (i.e., $\mathbf{z}_{1:q}$). Such transformations have a lower triangular Jacobian whose determinant is just the product of the diagonal elements, allowing us to create faster normalizing flows. The output vector $\mathbf{o}$ of a coupling flow is given by
\begin{align}\label{eq:cplng_lyr}
\mathbf{o}_{1:q} &= \mathbf{z}_{1:q} \nonumber \\
\mathbf{o}_{q+1:Q} &= \mathbf{g}_{({\mathbf{z}_{1:q}})}(\mathbf{z}_{q+1:Q}),
\end{align}
where $\mathbf{g}_{({\mathbf{z}_{1:q}})}$ is an invertible, element-wise transformation whose internal parameters have been computed based on $\mathbf{z}_{1:q}$ and in our case (conditional flows) also on the observed data $\mathbf{x}$. The final output of the transformation is then $\mathbf{o}=[\mathbf{o}_{1:q}, \mathbf{o}_{q+1:Q}]$. As coupling layers leave unmodified $\mathbf{z}_{1:q}$, one needs to shuffle the order of the input in each step using a permutation layer so that these two halves do not remain independent throughout the network.

For the coupling transformation $\mathbf{g}$, we have chosen a family of very expressive functions based on monotonically increasing splines \citep{Muller2018, Durkan2019}. They have demonstrated high flexibility when modeling multi-modal or quasi-discontinuous densities. A spline is a piece-wise function that is specified by the value at some key points called knots. 
The location, value, and derivative of the spline at the knots for each dimension in $\mathbf{o}_{q+1:Q}$ are calculated with a neural network. Each resulting distribution (and therefore each transformation) will depend on the observed data, and so the neural network will have the input $[\mathbf{z}_{1:q},\mathbf{x}]$.

\section{Hinode/SP dataset. Using Stokes $V$ information}\label{sec:appendix_hinode_stokesv}

In this section, we present the results of applying \Inspectorch\ to the Hinode/SP dataset using Stokes $V$ information, as a comparison to the results presented in the main text using only Stokes~$I$ information in Sect.~\ref{sec:hinode_case}. The Stokes~$V$ profiles are particularly useful for identifying magnetic features and flows in the solar atmosphere, as they show circular polarization signals at wavelengths far from the line core center only when there are strong flows, otherwise they are expected to be zero.

\begin{figure}[t]
    \centering
    \includegraphics[width=\linewidth]{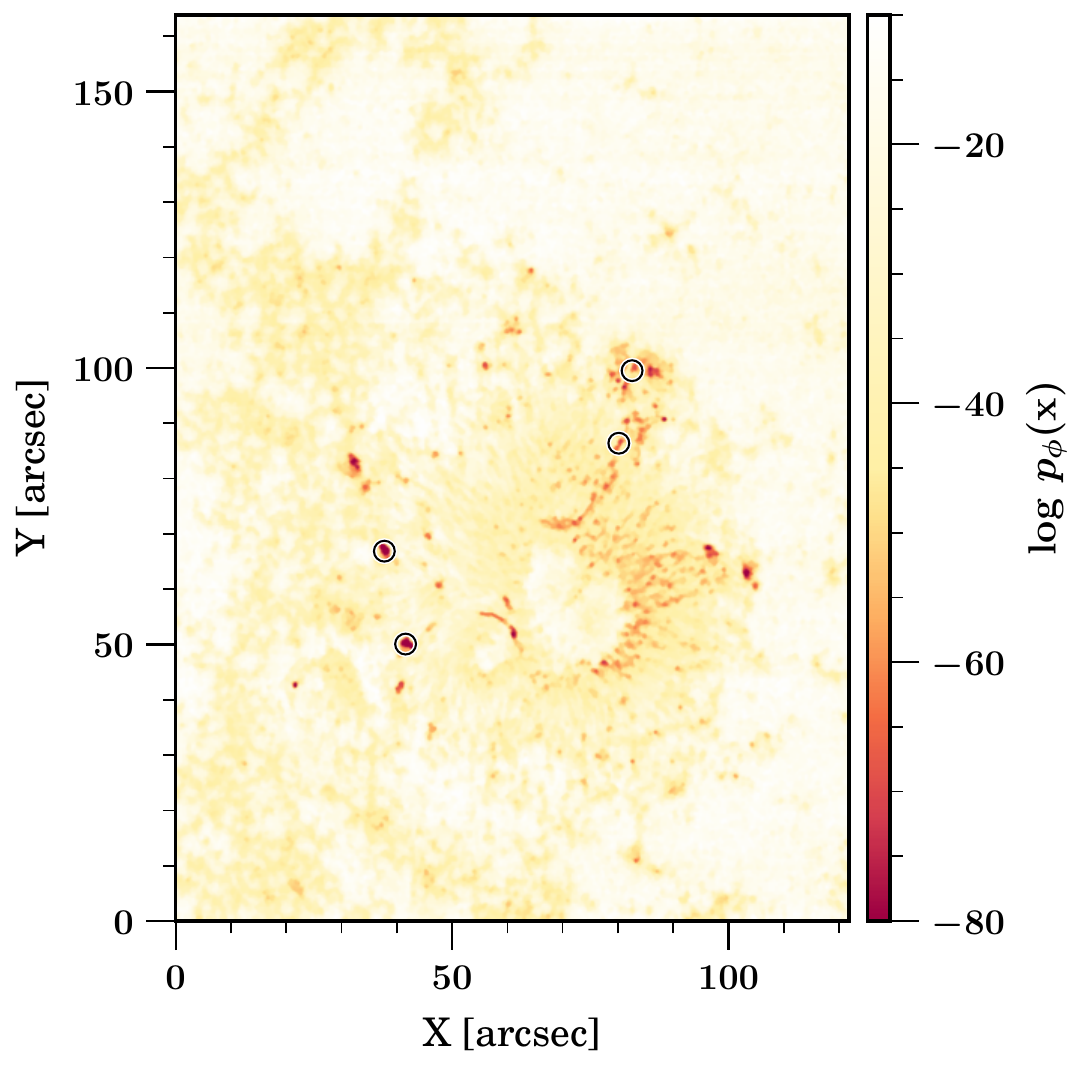}
    \caption{Log-probability map for the Hinode/SP dataset using Stokes $V$ information. The color scale in the log-probability map is clipped to a minimum to enhance contrast. Circled in black are the locations of four unusual profiles shown in Fig.~\ref{fig:hinode_4rare_spectra_stokesv}.}
    \label{fig:hinode_logprob_stokesv}
\end{figure}

\begin{figure}[t]
    \centering
    \includegraphics[width=0.95\linewidth]{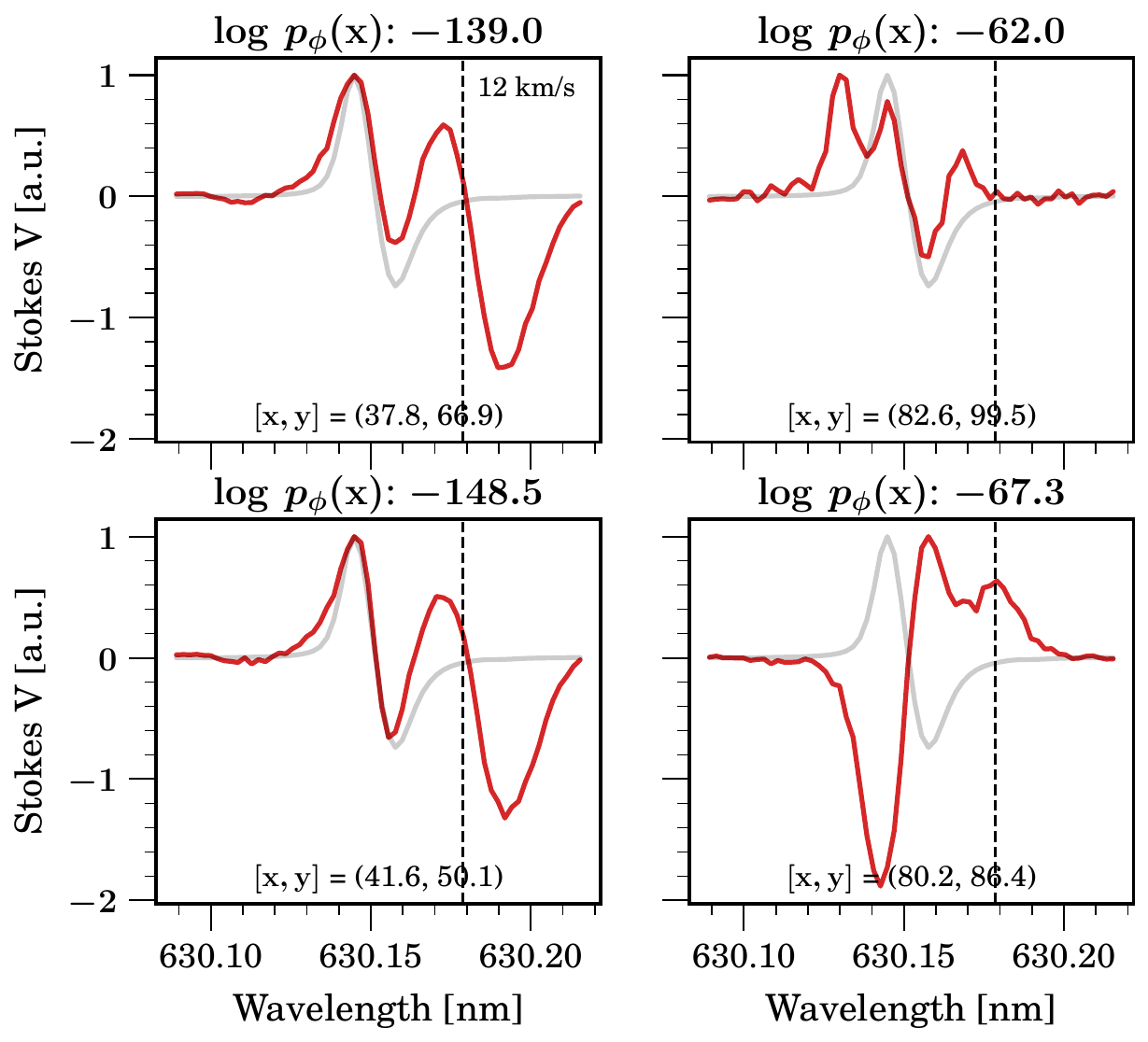}
    \caption{Four of the most unusual spectra identified in the Hinode/SP dataset using Stokes $V$ information. The spectra are normalized to their maximum value. The gray profile represents a typical circular polarization profile in this dataset. The dashed vertical line indicates a Doppler shift of 12~\kms. \fixii{The profile coordinates (in arcseconds) at the bottom of each panel correspond to the circled locations in Fig.~\ref{fig:hinode_logprob_stokesv}.}}
    \label{fig:hinode_4rare_spectra_stokesv}
\end{figure}

\end{document}